\documentclass[12pt]{article}
\usepackage{amsmath,amsfonts}
\usepackage{amsthm}
\usepackage{array}
\usepackage{textcomp}
\usepackage{stfloats}
\usepackage{url}
\usepackage{verbatim}
\usepackage{graphicx}
\usepackage{cite}
\usepackage[linesnumbered,ruled,vlined]{algorithm2e}
\usepackage[caption=false,font=normalsize,labelfont=sf,textfont=sf]{subfig}
\usepackage{makecell}
\usepackage{multirow}
\usepackage{comment}
\usepackage{enumitem}
\usepackage[normalem]{ulem}
\usepackage{pifont}
\usepackage{float}
\usepackage{booktabs}
\renewcommand\theadfont{\bfseries}
\SetKwInput{KwInput}{Input}
\SetKwInput{KwOutput}{Output}

\newtheorem{theorem}{Theorem}
\newtheorem{proposition}[theorem]{Proposition}
\newtheorem{protocol}{Protocol}

\newtheorem{algorisme}{Algorithm}
\usepackage[colorlinks=true, linkcolor=blue, citecolor=blue, urlcolor=blue,
            allcolors=blue]{hyperref}
\usepackage{authblk}

\begin{document}

\title{Efficient Unlearning with Privacy Guarantees}

\author[1]{Josep~Domingo-Ferrer}
\author[1]{Najeeb~Jebreel}
\author[1]{David~S\'anchez}
\affil[1]{CYBERCAT (Center for Cybersecurity Research of Catalonia), 
ComSCIAM-Center for Computational Science and Applied Mathematics, 
Department of Computer Engineering and Mathematics, 
Universitat Rovira i Virgili, Av. Pa\"{\i}sos Catalans 26, E-43007 
Tarragona, Catalonia \\
\texttt{\{josep.domingo, najeeb.jebreel, david.sanchez\}@urv.cat}}

\maketitle

\begin{abstract}
Legal frameworks recognize the right to request the forgetting of individuals' personal data.
In this respect, machine unlearning (MU) has emerged as a practical means to enable
digital forgetting by machine learning (ML) models trained on such data. 
However, existing MU approaches face trade-offs between utility of the ML model on 
retained tasks, forgetting guarantees, computational cost, and applicability to various ML models: 
exact unlearning methods are typically expensive, while more efficient approximate 
methods provide only empirical forgetting evidence, and certified methods offer guarantees inspired by
differential privacy (DP) 
only to the model distribution and often rely on restrictive assumptions about the model type or loss function.
In this paper, we present \emph{efficient unlearning with privacy guarantees} (EUPG), 
a novel framework that is {\em agnostic to the ML model type} and that {\em protects the data to be forgotten against disclosure with any privacy guarantee immune to post-processing} (including DP and some forms of $k$-anonymity).
EUPG involves training the ML model on data pre-protected under the selected
privacy guarantee and enables efficient unlearning through targeted fine-tuning.
We instantiate EUPG with probabilistic $k$-anonymity and $\epsilon$-DP, and evaluate it on four data sets with neural network and XGBoost models.
The results show that EUPG achieves the best trade-off between utility, efficiency, and unlearning effectiveness versus guaranteed unlearning baselines. 
Compared to exact unlearning, EUPG incurs substantially lower computational and storage costs; 
compared to certified unlearning, it applies to a wider range of ML model types and privacy guarantees,
and enforces protection at the training data level, which is more transparent and robust than guarantees defined 
on model parameters. 
\end{abstract}

\noindent\textbf{Keywords:} Machine unlearning, privacy, differential privacy, probabilistic $k$-anonymity, right to be forgotten.

\section{Introduction}
\label{sec:introduction}

Due to privacy concerns, individuals can request the removal of their personal data commonly used to train machine learning (ML) models.
The legal basis for these requests is the right to be forgotten, which is recognized by several data protection regulations, such as the European Union's General Data Protection Regulation (GDPR~\cite{GDPR}), the California Consumer Privacy Act (CCPA~\cite{CCPA}) and the recently approved European AI Act~\cite{euaiact}.
In particular, this right remains valid even if the user initially consented to the use of their data for ML. 

Since trained models may memorize and leak training data~\cite{geiping2020inverting,carlini2021extracting,balle2022reconstructing,wang2023reconstructing,yu2023bag,shokri2017membership,carlini2022membership}, 
they must be modified to forget the target data. 
A naive strategy for forgetting is to retrain the model from scratch after excluding the items whose removal has been requested.
However, retraining is often prohibitively expensive, especially with large-scale models and repeated deletion requests.

Machine unlearning (MU)~\cite{cao2015towards,bourtoule2021machine} aims to provide a cheaper alternative while maintaining the utility of the model on the retained tasks.
\emph{Exact unlearning} methods aim to obtain a model that behaves as if it had never seen the target records during training~\cite{bourtoule2021machine,yan2022arcane}.
Although this provides a perfect forgetting guarantee, it often entails substantial computational and storage costs, sometimes exceeding full retraining.
On the other hand, \emph{approximate unlearning} methods apply efficient \emph{ex post} updates to the trained model to mitigate the influence of the target data,
but provide only empirical evidence of forgetting~\cite{graves2021amnesiac,baumhauer2022machine,kurmanji2023towards,tarun2024fast}. Finally, \emph{certified unlearning} methods provide 
guarantees of indistinguishability inspired by differential privacy (DP~\cite{dwork2006differential}) 
between the model after unlearning and a model not having seen the target data~\cite{guo2020certified,sekhari2021remember,zhang2024towards,chien2024langevin,chien2024certified,qiao2025hessianfree,koloskova2025certified}.
Certified methods typically rely on injecting noise into model parameters 
during training, which we call in-protection; for this reason, they often require restrictive assumptions on convexity and/or smoothness of the model or the loss function.

Whereas exact, approximate, and certified unlearning seek to remove the influence of target data from the trained model, compliance with legal frameworks does not
require that much: specifically, anonymizing personally identifiable information (rather than deleting it) turns it into non-personal information
according to the GDPR and similar regulations and hence satisfies the right to be forgotten enshrined in those regulations.
This motivates a different perspective on unlearning: rather than relying only on \emph{ex post} modifications of the model parameters, one may instead protect training data \emph{before} training, so that post-unlearning releases provide unequivocal privacy guarantees for the data to be forgotten.

Based on this motivation, we propose a novel framework for machine unlearning, 
named \emph{efficient unlearning with privacy guarantees} (EUPG). 
EUPG adopts a \emph{data pre-protection} approach to provide formal privacy guarantees on the data that might be requested for unlearning. 
Specifically, before model training, 
we protect
the training data with a privacy guarantee immune to post-processing,
which will be preserved by the output of subsequent training or fine-tuning processes. We then train a base model on the protected data to get a protected model. To recover utility, this protected model is fine-tuned
on the original training data,
which yields a competitive (but unprotected) production model. Upon a forgetting request, we roll back to 
the protected model and 
efficiently fine-tune it {\em only on the original training data to be retained}. In this way, the model after unlearning acquires the pre-enforced privacy guarantee {\em on the data to be forgotten}.
Compared with exact unlearning, EUPG substantially reduces computational and storage costs. 
Unlike certified unlearning methods whose guarantees are stated at the model level, 
EUPG grounds its guarantee at the training data level, and thus {\em works with any ML model}: since protection is applied before training, 
the privacy guarantee is tied to the protected training data rather than the parameter space of the model. 
In fact, \cite{thudi2022necessity} argue that matching or approximating a retrained model in the parameter space may fail to provide sound evidence of forgetting. 
Thus, EUPG avoids several limitations of certified unlearning methods, including restrictive assumptions on the model or the loss function, 
and remains 
model-agnostic, because its guarantee does not depend on the optimization dynamics of the downstream model. 
Finally, EUPG can accommodate a broad class of privacy guarantees
that are only required to be immune to post-processing (this class includes but is not limited to DP). 

In summary, our contributions are:
\begin{itemize}
    \item An efficient machine unlearning framework that provides formal privacy guarantees on the data to be forgotten. The
    framework is agnostic to the ML model and the loss function,
    and in particular to their convexity or smoothness.

    \item A first instantiation of our framework with probabilistic $k$-anonymity~\cite{soria2012probabilistic} as a privacy guarantee. 
    The resulting method protects the unlearned data under probabilistic $k$-anonymity. 

    \item A second instantiation with differential privacy (DP) \cite{dwork2006differential} as a privacy guarantee. The resulting method protects the unlearned data under $\epsilon$-DP. 

    \item Experiments and detailed analyses on four data sets, involving tabular data and images, showing that EUPG achieves competitive model utility and forgetting effectiveness relative to exact and certified unlearning baselines. 
\end{itemize}

The remainder of this paper is organized as follows.
Section~\ref{sec:related_work} reviews related work on machine unlearning.
Section~\ref{sec:background} provides background on the privacy guarantees used in our instantiations.
Section~\ref{sec:eupg} presents the EUPG framework and formally demonstrates its guarantees.
Section~\ref{sec:experimental} reports experimental results on a variety of data sets and ML models.
Section~\ref{sec:conclusion} contains conclusions and directions for future work.

\section{Related work}
\label{sec:related_work}

Machine unlearning seeks to remove the influence of a subset of training data from a learned model, 
while avoiding the cost of full retraining.
Depending on the kind of guarantees on the unlearned data items they offer, 
MU methods can be classified as exact,
approximate, or certified~\cite{nguyen2025survey,xu2024machine,blanco2025digital}.

\paragraph{Exact unlearning.}
Exact unlearning seeks to completely remove the influence
of the unlearned data, ideally matching retraining from scratch on the retained data.
SISA~\cite{bourtoule2021machine} is the best-known exact method. It partitions training data into shards and slices, stores intermediate checkpoints during training, and retrains only the affected shard(s) starting from the last checkpoint preceding the slice containing the data to be forgotten.
Although it achieves exact unlearning, it incurs high storage and computational overhead due to checkpointing and ensemble inference, and it may hamper model utility because each shard is trained on only a small subset of the data.
Other exact approaches exist for specific model types, including clustering~\cite{ginart2019making}, tree ensembles~\cite{schelter2021hedgecut,brophy2021machine}, and graph models~\cite{chen2022graph}.

\paragraph{Approximate unlearning.}
Approximate unlearning methods apply efficient \emph{ex post} updates that empirically reduce the influence of the unlearned data.
Representative strategies include fine-tuning using the retained data and gradient ascent~\cite{golatkar2020eternal}, 
teacher-student objectives~\cite{kurmanji2023towards}, 
class-level impair-repair procedures~\cite{tarun2024fast}, 
and methods that fine-tune the model on a randomly relabeled forget set or subtract its stored per-batch gradient updates from the learned parameters~\cite{graves2021amnesiac}.
Although computationally efficient, these methods do not provide formal unlearning guarantees 
and thus fall short of legal compliance with the right to be forgotten.

\paragraph{Certified unlearning.}
Certified unlearning provides formal DP-style guarantees that bound the distinguishability between the distribution of models produced by an unlearning algorithm and that of 
models that have never been trained on the unlearned data.
This notion, commonly referred to as $(\epsilon,\delta)$-unlearning, formalizes unlearning as a distribution-level indistinguishability requirement~\cite{guo2020certified,sekhari2021remember}.
Certified unlearning may thus be viewed as approximate unlearning with a formal probabilistic guarantee (bounded by the chosen parameters), rather than exact influence removal.
Since these methods typically rely on injecting noise into model parameters, much of the prior work establishes certified guarantees primarily under convexity or strong convexity assumptions~\cite{guo2020certified,neel2021descent,sekhari2021remember,warnecke2023machine}, thus limiting the applicability to neural networks.
Recent attempts to extend certification to non-convex settings often impose additional assumptions such as smoothness, bounded gradients, or calibrated perturbation tied to optimization constants~\cite{chien2024langevin,chien2024certified,mu2025rewindtodelete,qiao2025hessianfree,zhang2024towards}.
In particular, Rewind-to-Delete~\cite{mu2025rewindtodelete} provides certified unlearning for general non-convex optimization, but still relies on smoothness-type assumptions.
More recently, \cite{koloskova2025certified} have proposed certified unlearning for neural networks via privacy amplification by stochastic post-processing, 
using noisy fine-tuning on retained data with clipping.
A key advantage is that their guarantees are derived without explicitly requiring smoothness assumptions on the loss function, which improves applicability to practical deep learning settings.
However, since certification is enforced through clipping and calibrated Gaussian perturbation in the parameter space, utility can degrade noticeably as model dimensionality increases.
Furthermore, achieving a favorable utility--privacy tradeoff requires jointly tuning several interacting unlearning-phase hyperparameters whose optimal combination is data set- and model-specific and typically requires dedicated hyperparameter sweeps.
Finally, because these methods operate at the model parameter level, their guarantees are generally less precise and harder to interpret than those of data-level approaches, such as exact unlearning. 
In fact, \cite{thudi2022necessity} {\em object to parameter-space distinguishability criteria as evidence of forgetting}, arguing that {\em similar or identical model parameters may be reachable from different training data sets}.

\paragraph{Our framework.}
Unlike certified unlearning methods, which place the guarantee at the \emph{model level}, 
EUPG places it at the \emph{data level}, 
which is more precise and transparent: the only channel through which forgotten records can influence any post-unlearning release is the protected data set generated before training.
Unlike approximate unlearning methods, which provide only empirical removal, our approach offers a formal post-unlearning privacy guarantee for the requested forget set.
Unlike exact approaches, such as retraining or SISA, EUPG can satisfy unlearning requests without repeatedly retraining from scratch or maintaining multiple checkpoints, substantially reducing operational costs.

Moreover, our framework is agnostic to the downstream ML model. In 
fact, once a protected training data set has been produced, the post-unlearning privacy guarantee, being
immune to post-processing, is inherited despite subsequent training, rollback, and fine-tuning on retained data.
Furthermore, our framework requires minimal unlearning-specific hyperparameter selection: the post-unlearning step reduces to standard fine-tuning on 
retained data, with no additional clipping thresholds, noise schedules, or regularization coefficients to be jointly tuned (in contrast to certified perturbation-based methods like~\cite{koloskova2025certified}).
Overall, this design decouples the post-unlearning privacy guarantee from the architecture of the downstream ML model and enables efficient plug-and-play deployment across a broad range of learning algorithms.

Table~\ref{tab:unlearning_comparison} provides a comparative overview of the main unlearning paradigms and our approach, highlighting their guarantee level, applicability, and operational trade-offs.

\begin{table*}[t!]
\centering
\caption{Overview of machine unlearning paradigms and their guarantees}
\label{tab:unlearning_comparison}
\renewcommand\theadfont{\bfseries}
\setlength{\tabcolsep}{6pt}
\resizebox{\linewidth}{!}{%
\begin{tabular}{lccccccc}
\toprule
\thead{Paradigm} &
\thead{Representative\\methods} &
\thead{Guarantee\\level} &
\thead{Non-convex\\ applicability} &
\thead{No smoothness\\ restrictions} &
\thead{Model\\agnostic} &
\thead{Privacy\\model flexibility} &
\thead{Unlearning\\cost} \\
\midrule

\multirow{2}{*}{Exact unlearning}
  & Retrain from scratch
  & \multirow{2}{*}{\shortstack{Data-level \\ (exact data removal)}}
  & Yes & Yes & Yes & N/A & Significant \\
  & SISA~\cite{bourtoule2021machine}
  &  & Yes & Yes & Yes & N/A & Significant/High \\
\midrule

Approximate unlearning
  & \cite{graves2021amnesiac,kurmanji2023towards,tarun2024fast}
  & \shortstack{Model-level \\ (empirical forgetting)}
  & Yes & Yes & No & N/A & Low \\
\midrule

\multirow{4}{*}{\makecell[l]{Certified\\$(\epsilon,\delta)$ unlearning}}
  & Convex / strongly-convex methods~\cite{guo2020certified,neel2021descent,sekhari2021remember,warnecke2023machine}
  & \multirow{4}{*}{\shortstack{Model-level \\ (distribution indistinguishability)}}
  & No & No & No & No & Moderate/High \\
  & Noisy gradients~\cite{chien2024langevin,chien2024certified}
  &  & Partial & No & No & No & Moderate \\
  & Rewind-to-Delete~\cite{mu2025rewindtodelete}
  &  & Yes & No & No & No & Moderate \\
  & Stochastic post-processing~\cite{koloskova2025certified}
  &  & Yes & Yes & No & No & Low \\
\midrule

Training data pre-protection
  & EUPG
  & \shortstack{Data-level \\ (privacy guarantee)}
  & Yes & Yes & Yes & Broad$^\dagger$ & Low \\
\bottomrule
\end{tabular}}
\vspace{1mm}

{\footnotesize $^\dagger$Supports any post-processing immune privacy guarantee ({\em e.g.}, differential privacy and probabilistic $k$-anonymity).}
\end{table*}

\section{Background}
\label{sec:background}

This section provides background on data anonymization and the main two families of privacy guarantees (also known
as privacy models).

Tabular data sets, also called microdata sets, are often used to train ML models. 
A tabular data set is composed of records, where each record reports several attributes on an individual. 
Considering their disclosure potential, attributes can play the following roles: (i) {\em identifiers} ({\em e.g.}, passport no., social security no., name-surname, etc.) unequivocally identify the individual to whom
the record corresponds; (ii) {\em quasi-identifiers} ({\em e.g.}, zipcode, profession, gender, age, etc.) do not identify the individual when taken separately,
but they may when considered jointly ({\em e.g.}, a 90-year-old female doctor living in a rural zipcode is probably unique);
(iii) {\em confidential attributes} ({\em e.g.}, diagnosis, income, etc.) contain sensitive 
information; iv) other attributes may exist that are not in the previous three groups.

When data are anonymized, the identifiers must be suppressed.
However, this is not enough, and some modification of the quasi-identifiers is needed
to prevent {\em re-identification} by an adversary.
To offer \textit{ex ante} guarantees against re-identification, this modification should be done under the scope of a privacy guarantee.

$k$-Anonymity~\cite{samarati2001protecting} was the first
privacy guarantee proposed, and it was followed by several
variants and extensions (like $l$-diversity~\cite{ldiversity} or $t$-closeness~\cite{tcloseness},
which also protect against disclosure of confidential attributes). A data set is $k$-anonymous if, for each
combination of quasi-identifiers present in the data set, 
there are at least $k$ records sharing that combination.
These records form a so-called $k$-anonymous class. 
In this way, the probability
of successful re-identification is at most $1/k$.
A data set can be made $k$-anonymous using generalization 
and suppression of {\em quasi-identifiers} (the approach of
the original article~\cite{samarati2001protecting})
or using microaggregation~\cite{domingo2005ordinal}.
The latter method
microaggregates records by their quasi-identifier attributes, that is, creates clusters of at least $k$ records
(and less than $2k$ records) such that the
quasi-identifier values within each cluster are maximally similar. 
Then, the centroid of the quasi-identifier values in each cluster
is computed and used to replace the quasi-identifier values
of the records in the cluster. 
With all methods, confidential attributes are left unchanged.

The above $k$-anonymity definition has several issues. First, the need to decide
which attributes are quasi-identifiers has been pointed out as a weakness;
however, the data protector can circumvent this by taking any attribute as a quasi-identifier unless they are completely
sure it cannot be used as such by an attacker. 
More serious are the weaknesses due to the syntactic
nature of $k$-anonymity. Specifically, the need to ensure that each combination of 
quasi-identifiers appears at least $k$ times in the anonymized file 
precludes post-processing immunity: if post-processing a $k$-anonymous
file makes a combination appear less than $k$ times, we no longer have
$k$-anonymity. Also, as pointed out in~\cite{cohen}, if $k$-anonymity
is enforced deterministically via minimum generalizations and these are known,
in some cases $k$-anonymity may be partially reversible.

A non-syntactic, that is, a semantic version of $k$-anonymity can bypass the above shortcomings, as it is the case for {\em probabilistic $k$-anonymity}~\cite{soria2012probabilistic}.
A data set $\mathcal{D}'$ generated from an original data set $\mathcal{D}$ via mechanism $\mathcal{A}$ is said to satisfy {\em probabilistic $k$-anonymity} if, for any non-anonymous external data set $\mathcal{E}$, the probability that an intruder knowing $\mathcal{D}'$, $\mathcal{A}$ and $\mathcal{E}$ correctly links any record in $\mathcal{E}$ with its corresponding record (if any) in $\mathcal{D}'$ is at most $1/k$.

The algorithm used 
in~\cite{soria2012probabilistic} to achieve probabilistic
$k$-anonymity is based
on randomly permuting quasi-identifier combinations within
clusters of at least $k$ records. 
These clusters are computed using microaggregation~\cite{domingo2005ordinal}, but instead 
of replacing the quasi-identifiers in each cluster by
their centroid, 
quasi-identifier combinations are randomly permuted within 
the cluster, following 
the Anatomy approach~\cite{xiao2006anatomy}.  
This stochastic approach ensures 
irreversibility, that is, the original data cannot be recovered
from the probabilistically $k$-anonymous data. 
Note that probabilistic $k$-anonymity is no longer based 
on a syntactic condition such as a $k$-fold repetition of quasi-identifier combinations: it just focuses on the semantic
requirement that the risk of re-identification be at most $1/k$.
When achieved stochastically, probabilistic
$k$-anonymity offers post-processing immunity: post-processing
cannot undo the random permutation that ensures that the risk
is at most $1/k$ 
\cite{models}.

$\epsilon$-Differential privacy (DP, \cite{dwork2006differential}) defines a more stringent privacy guarantee that subsequently inspired several variants or relaxations~\cite{dwork_foundations}. DP seeks to bound the influence of any particular record in the original data set (and thus of the individual to whom the record corresponds) on the statistical outcomes obtained from the data set. Formally speaking, a randomized query
function $\kappa$ satisfies $\epsilon$-differential privacy if, for all data sets $D_1$ and $D_2$ that differ in one
record (also known as neighbor data sets), and all $S \subset Range(\kappa)$, we have $\Pr(\kappa(D_1) \in S) \leq \exp(\epsilon) \Pr(\kappa(D_2) \in S).$

For a numerical query $f$, $\epsilon$-DP can be achieved through noise addition, that is, by computing $\kappa(x)=f(x)+N$, 
where $N$ stands for noise.
Typically, the Laplace noise distribution is used. The amount of noise that must be added is inversely proportional to $\epsilon$ (smaller $\epsilon$ means greater privacy and thus requires more noise) and directly
proportional to the sensitivity of the query function $f$ (variability between neighbor data sets), that is, 
$\Delta_f= \max_{D_1,D_2 \in {\mathcal D}} ||f(D_1) - f(D_2)||_1$,
where $D_1$ and $D_2$ are two data sets that differ in one record, and $\mathcal D$ is the collection of data sets on which $f$ can be evaluated.
For non-numerical queries, the exponential mechanism~\cite{mcsherry2007mechanism} is typically used. This mechanism selects outputs with a probability proportional 
to an exponential function of the utility of the outputs, ensuring that more ``useful'' results are more likely, while still protecting privacy.

DP has some interesting properties:
\begin{itemize}
    \item {\em Immunity to post-processing}. If a function $f$ provides $\epsilon$-DP, then any function $g$, applied to the output of $f$, also preserves $\epsilon$-DP.
    \item {\em Sequential composition}.
 Let $\kappa_{1}$ be a randomized function that satisfies $\epsilon_{1}$-DP
	and $\kappa_{2}$ a randomized function that satisfies $\epsilon_{2}$-DP.
	Then, any deterministic function of $(\kappa_{1},\kappa_{2})$ satisfies
	$(\epsilon_{1}+\epsilon_{2})$-DP. 

\item {\em Parallel composition}. Let $\kappa_{1}$ and $\kappa_{2}$ be randomized functions
that satisfy $\epsilon$-DP. If $\kappa_{1}$ and $\kappa_{2}$ are applied
	to {\em disjoint} data sets, any deterministic
	function of $(\kappa_{1},\kappa_{2})$ satisfies $\epsilon$-DP. 
\end{itemize}

Although DP was initially proposed to protect queries against a remote database, it can also be applied to anonymize tabular microdata sets by enforcing the Laplacian or exponential mechanisms on \emph{all} attributes of the microdata set \cite{soria2014enhancing, sanchez2016individualranking}. 
In this case, the global sensitivity is the maximum variability of each attribute (typically, the attribute range). 
Moreover, since attribute values within a record are usually correlated, sequential composition applies:
hence, the budget $\epsilon$ needs to be split among the attributes in the data set, which means that the noise added to each attribute increases with the number of attributes.

\section{Efficient Unlearning with Privacy Guarantees}
\label{sec:eupg}

\begin{figure*}[t!]
    \centering
    \includegraphics[width=\linewidth]{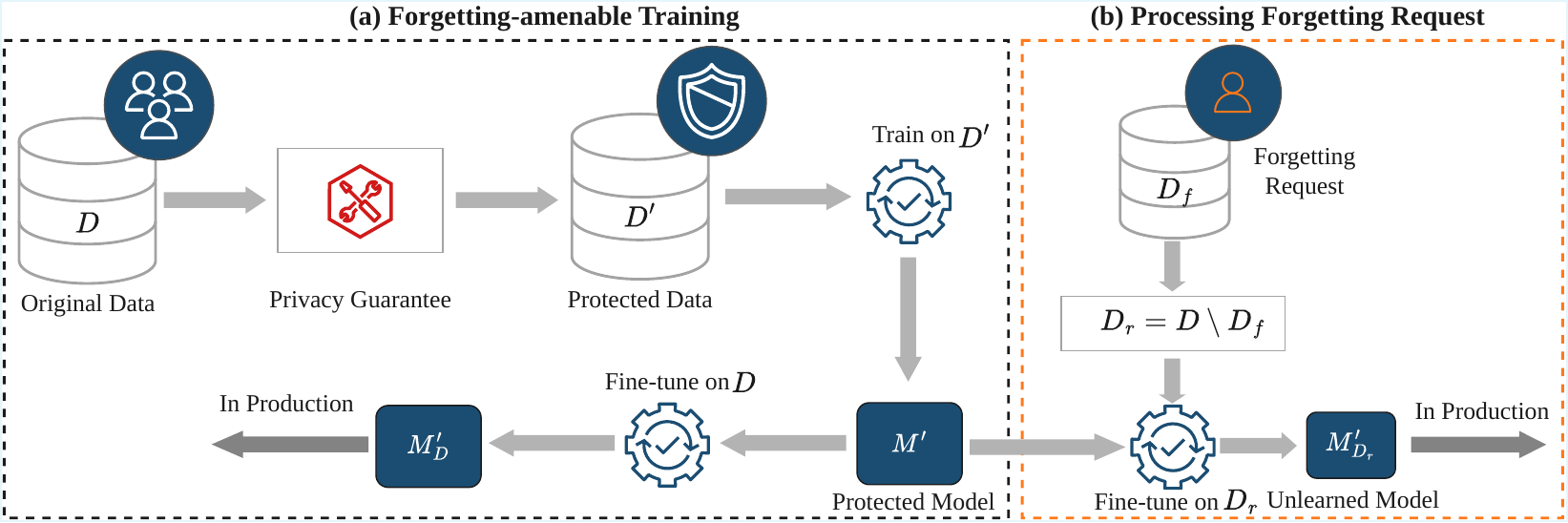}
    \caption{Workflow of the proposed EUPG framework}
    \label{fig:EUPG_pipeline}
\end{figure*}

We introduce \emph{efficient unlearning with privacy guarantees} (EUPG), a framework designed to simultaneously address flexibility, privacy, utility, and efficiency in unlearning. 
The core idea is to shift the source of the formal guarantee to a randomized protection mechanism applied 
to the training data \emph{before} model training, and then to ensure that every post-unlearning release is obtained only from the protected training data together with the {\em retained} original training data.
In this way, EUPG avoids repeated retraining from scratch and the limitations of existing in-protection certified mechanisms while still providing a formal privacy guarantee for the unlearned data.
Figure~\ref{fig:EUPG_pipeline} illustrates the EUPG workflow, which consists of two conceptual stages.

\paragraph{Stage I: Forgetting-amenable training} 
This stage involves preparing the ML model in a way that inherently supports efficient and effective unlearning. 
First, a post-processing immune privacy guarantee is chosen 
(such as those introduced in Section \ref{sec:background} or a variant/extension of them). 
The choice of the privacy guarantee depends on multiple factors, including the types of attributes and the motivation to unlearn.
For a particular privacy guarantee, a stronger privacy level requires stricter parameters ({\em e.g.}, larger $k$ for probabilistic 
$k$-anonymity or
smaller $\epsilon$ for DP).

Once a privacy guarantee is chosen, the original data set ($\mathbf{D}$) is transformed into a protected version ($\mathbf{D'}$) by a mechanism that enforces the privacy guarantee (as described in Section \ref{sec:background}). 
An initial ML model $M'$ is then pre-trained on $\mathbf{D'}$. Since the training procedure accesses only the protected data $\mathbf{D'}$, the resulting model $M'$ inherits the privacy guarantee of the protection mechanism by its post-processing immunity. In this sense, protection is established at the data level and therefore extends to any downstream processing of $\mathbf{D'}$, not only to model training.
This initial data protection and the pre-training steps are the most computationally expensive part of our framework, but they are one-time processes that set the stage for efficient, utility-preserving unlearning with privacy guarantees.

Since enforcing data protection for $\mathbf{D'}$ necessarily entails a loss of information, the utility of $M'$ is likely to degrade, which can limit model competitiveness in production environments. 
In particular, privacy-preserving perturbations will cause $M'$ to learn only general patterns without memorizing consistent item-level information.
To mitigate utility degradation, we fine-tune $M'$ on the entire original data set $\mathbf{D}$, which allows $M'$ to quickly make up for the lost utility: now the model is rapidly learning only the item-level information it missed, rather than general patterns that are slower to consolidate. 
The resulting fine-tuned model $M'_{\mathbf{D}}$ is then deployed for production. Notice that this first production model does not offer any protection, as it has been fine-tuned on the entire original data.

\paragraph{Stage II: Processing of forgetting requests}
Suppose that a forgetting request specifies a forget set $\mathbf{D}_f \subseteq \mathbf{D}$,
and let the retained set be $\mathbf{D}_r = \mathbf{D} \setminus \mathbf{D}_f$.
Rather than retraining from scratch on $\mathbf{D}_r$, EUPG discards the currently
deployed model $M'_{\mathbf{D}}$, rolls back to the protected base model $M'$, and
fine-tunes $M'$ exclusively on $\mathbf{D}_r$.
The resulting model $M'_{\mathbf{D}_r}$ is then released as the new post-unlearning model.

This design has important consequences.
First, since raw items of $\mathbf{D}_f$ are never used after rollback, the released
model $M'_{\mathbf{D}_r}$ depends on the forget set only through the already protected
model $M'$ produced before training, whose privacy guarantees are inherited
from the pre-training protection phase.
Hence, the privacy of individuals whose data reside in $\mathbf{D}_f$ is protected
\emph{by design} under the privacy guarantee embedded in $M'$.

Second, since fine-tuning starts from a trained base model $M'$ rather than from a random
initialization, the computational cost of satisfying each unlearning request will be much lower than that of exact unlearning via retraining.
The number of post-unlearning fine-tuning epochs then becomes an explicit 
control knob for the utility--efficiency trade-off: 
a model manager who prioritizes utility (resp. low cost) can increase (resp. decrease) the number of fine-tuning epochs, 
while the formal privacy guarantee is unaffected by this choice and remains tied to the original protection mechanism.

\paragraph{Threat model.}
We consider an adversary whose goal is to infer membership in $\mathbf{D}_f$
or to reconstruct information about forgotten records, by observing the
post-unlearning model $M'_{\mathbf{D}_r}$ ---or querying it--- in a white-box or black-box setting.
Access to the pre-unlearning model $M'_{\mathbf{D}}$ or any intermediate
artifact is not assumed; what an adversary may have inferred before the
unlearning request is outside the scope of EUPG, as is the case for most certified and exact unlearning
approaches, including retraining from scratch \cite{mdai2025}.
Also, it is the model manager's responsibility to make sure that \emph{all} of the requesting individual's data are forgotten, that is, no dependencies between the forgotten data ${\bf D}_f$ and the 
retained data ${\bf D}_r$ remain. In any case, we assume that the adversary does not know of any dependencies between ${\bf D}_f$ and ${\bf D}_r$ that they could exploit to make inferences on the former based on the latter. This assumption, which we shall call ``no known dependencies between forgotten and retained data'', pervades (often implicitly) all the literature on machine unlearning and data forgetting. 

In the following, we show how this general framework can be instantiated with the two families of privacy guarantees introduced in Section \ref{sec:background}, and we prove the resulting unlearning guarantees with respect to the threat model.

\subsection{Unlearning with a probabilistic $k$-anonymity privacy guarantee}
\label{sec:k_anonym_mu}

Under probabilistic $k$-anonymity, the probability that an adversary can re-identify a forgotten
record after unlearning 
based on the quasi-identifiers 
is upper bounded by $1/k$.

\begin{algorisme}[Probabilistic $k$-anonymity amenable training]
\label{alg1}~
\begin{enumerate}
\item Let $\mathbf{D}$ be a data set used for training.
Let $QI$ be the set of quasi-identifier attributes in $\mathbf{D}$.
\item Transform $\mathbf{D}$ into a probabilistically
$k$-anonymous $\mathbf{D}^k$ by 
grouping records into clusters of size at least $k$ and
applying 
a random permutation method 
on the $QI$ combinations within each cluster
(see Section~\ref{sec:background}). 
\item Train a machine learning model on $\mathbf{D}^k$. Let
the trained model be $M^k$.
\item\label{fine} Fine-tune $M^k$ on $\mathbf{D}$ to obtain
a model $M^k_\mathbf{D}$.
\end{enumerate}
\end{algorisme}

\begin{protocol}[Probabilistically $k$-anonymous unlearning]
\label{pro1}~
\begin{enumerate}
\item An individual asks the manager of model $M^k_\mathbf{D}$
to remove her data $\mathbf{D}_f$ from $\mathbf{D}$ and $M^k_\mathbf{D}$.
\item The model manager deletes $M^k_\mathbf{D}$ and fine-tunes $M^k$ on $\mathbf{D}_r = \mathbf{D} \setminus \mathbf{D}_f$ to obtain $M^k_{\mathbf{D}_r}$.
\end{enumerate}
\end{protocol}

\begin{proposition}[Privacy]
\label{prop1}
Under the assumption of no known dependencies between 
the forgotten and retained data, 
unlearning with Protocol~\ref{pro1} satisfies 
probabilistic $k$-anonymity,
that is, the individuals in the 
set $\mathbf{D}_f$ to be forgotten are protected by probabilistic $k$-anonymity.
\end{proposition}

{\bf Proof:} In Protocol~\ref{pro1}, the model manager reverts to  
$M^k$, which has been trained on the probabilistically $k$-anonymous
version $\mathbf{D}^k$ of the original data set $\mathbf{D}$.
Then, fine-tuning to obtain $M^{k}_{\mathbf{D}_r}$ excludes the forget set $\mathbf{D}_f$ and, under the assumption that there are no known dependencies, 
introduces no 
additional information 
on $\mathbf{D}_f$ into the fine-tuned model. 
According to the property of post-processing immunity of 
probabilistic $k$-anonymity (see Section~\ref{sec:background}), the 
probabilistic $k$-anonymity guarantee
achieved by $\mathbf{D}^k$ extends to $M^{k}_{\mathbf{D}_r}$
for the individual(s) to whom $\mathbf{D}_f$ corresponds. \hfill $\Box$

\subsection{Unlearning with a differentially private guarantee}
\label{sec:dp_mu}

We next show how to obtain an $\epsilon$-DP guarantee for the forgetting request, which means that, after unlearning, the forgotten data should be unnoticeable from the output of the ML model
except by a factor $\exp(\epsilon)$.

\begin{algorisme}[DP-amenable training]
\label{alg2}~
\begin{enumerate}
\item Let $\mathbf{D}$ be a data set used for training.
\item Transform $\mathbf{D}$ into an $\epsilon$-DP data set
$\mathbf{D}^\epsilon$ ({\em e.g.}, by applying the Laplacian or exponential mechanisms to every attribute; see Section~\ref{sec:background}). If record attributes are not independent and thus sequential composition applies, use a budget $\epsilon/|attributes|$ for each attribute to obtain $\epsilon$-DP protected records. 
\item Train a machine learning model on $\mathbf{D}^\epsilon$. Let
the trained model be $M^\epsilon$.
\item Fine-tune $M^\epsilon$ on $\mathbf{D}$ to obtain
a model $M^{\epsilon}_\mathbf{D}$.
\end{enumerate}
\end{algorisme}

\begin{protocol}[$\epsilon$-DP unlearning]
\label{pro2}~
\begin{enumerate}
\item An individual asks the manager of model $M^{\epsilon}_\mathbf{D}$
to remove her data $\mathbf{D}_f$ from $\mathbf{D}$ and $M^{\epsilon}_\mathbf{D}$.
\item The model manager fine-tunes $M^\epsilon$ on
$\mathbf{D}_r = \mathbf{D} \setminus \mathbf{D}_f$ to obtain $M^{\epsilon}_{\mathbf{D}_r}$.
\end{enumerate}
\end{protocol}

\begin{proposition}[Privacy]
\label{prop2}
Under the assumption of no known dependencies between 
the forgotten and retained data, 
unlearning with Protocol~\ref{pro2} satisfies
$\epsilon$-DP, that is, the individuals in the set
$\mathbf{D}_f$ to be forgotten are protected by $\epsilon$-DP.
\end{proposition}

{\bf Proof:} In Protocol~\ref{pro2}, the model manager reverts to
$M^\epsilon$, which has been trained on the $\epsilon$-DP
version $\mathbf{D}^\epsilon$ of the original data set $\mathbf{D}$.
Then, fine-tuning to obtain $M^{\epsilon}_{\mathbf{D}_r}$ excludes the forget set $\mathbf{D}_f$,
and, under the assumption that there are no known dependencies, introduces no 
additional information 
on $\mathbf{D}_f$ into the fine-tuned model. 
According to the property of post-processing immunity of DP (see Section~\ref{sec:background}), the $\epsilon$-DP guarantee achieved by $\mathbf{D}^\epsilon$ extends to $M^{\epsilon}_{\mathbf{D}_r}$
for the individual(s) to whom $\mathbf{D}_f$ corresponds.
\hfill $\Box$

$\epsilon$-DP can be replaced by any of its variants or
relaxations, and it is straightforward to adapt
Algorithm~\ref{alg2}, Protocol~\ref{pro2}, and Proposition~\ref{prop2}. 
Note that post-processing immunity holds for all state-of-the-art variants and relaxations of DP.

\section{Experimental results}
\label{sec:experimental}

We evaluated EUPG on four data sets and three models, measuring utility, forgetting effectiveness, and computational efficiency. 
We compared our approach with unlearning methods also providing forgetting guarantees: exact unlearning (retraining from scratch and SISA~\cite{bourtoule2021machine}) and a state-of-the-art certified method (Certified-SP) based on privacy amplification by stochastic post-processing~\cite{koloskova2025certified}.

Experiments were conducted on a Windows 11 system using Ubuntu 20.04 (WSL2), equipped with an Intel\textregistered Core\texttrademark i7-12700 CPU (12 cores), 32 GB RAM, and an NVIDIA RTX 4080 GPU (16 GB VRAM).
Our code is available at \url{https://github.com/najeebjebreel/eupg}

\subsection{Experimental setup}
\label{sec:setup}

\paragraph{Data sets} We used four publicly available data sets, each representing a classification problem from a different domain. 
Three of these data sets are tabular and were chosen because of their privacy relevance, as they contain records describing personal data of individuals. 
In addition, we included an image classification data set to evaluate the generality of our approach across various data types.
\begin{itemize}
\item \textit{Adult income~\footnote{\url{https://archive.ics.uci.edu/ml/datasets/Adult}}}: It comprises 32,561 training records and 16,281 testing records of demographic and financial data, with six numerical and eight categorical attributes. The class attribute indicates whether an individual makes more than 50K dollars a year. 
\item \textit{Heart disease~\footnote{\url{https://www.kaggle.com/sulianova/cardiovascular-disease-dataset}}}: It contains 55,869 training records and 14,131 testing records of patient data, with five numerical measurements and six categorical measurements related to cardiovascular diseases. The class attribute denotes the presence of heart disease.
\item \textit{Credit information~\footnote{\url{https://www.kaggle.com/c/GiveMeSomeCredit}}}: It includes 96,215 training records and 24,054 testing records of financial information, with ten numerical attributes. The class attribute indicates whether an individual has experienced financial distress.
\item \textit{CIFAR10 \footnote{\url{https://www.cs.toronto.edu/~kriz/cifar.html}}}:
It is a widely used image classification data set that contains 60,000 32x32 pixel images in ten classes, with three RGB channels. The data set is divided into 50,000 training samples and 10,000 testing samples.
\end{itemize}

\paragraph{ML models} For each tabular data set, we built two ML classification models: a multi-layer perceptron (MLP) and an XGBoost classifier. 
We implemented MLP models using PyTorch with an input layer, two hidden layers of equal size, and an output layer. 
Regarding XGBoost classifiers, we utilized the implementation provided by XGBoost~\cite{chen2016xgboost}, which can be found on the official XGBoost website~\footnote{\url{https://xgboost.ai/}}.
For CIFAR10, we used the ResNet18~\cite{he2016deep} deep model.

\paragraph{Evaluation metrics} The utility of the ML models was evaluated using accuracy (Acc) for the Adult, Heart, and CIFAR10 data sets. 
For the Credit data set, which has a significant class imbalance (only 6.92\% of the data belong to the positive class), the area under the ROC curve (AUC) was used instead.
The effectiveness of forgetting was assessed using RMIA~\cite{zarifzadeh2024lowcost}, 
a membership inference attack that scores each sample by comparing its likelihood under the target model with that of the reference models. 
We used 8 offline reference models for RMIA and evaluated it using both AUC and the true positive rate at 1\% false positive rate (TPR@1\%FPR), which are the two main metrics used to evaluate MIA effectiveness.

\paragraph{Privacy guarantees} 
To allow a fair comparison between probabilistic $k$-anonymity and DP, we applied both privacy guarantees to all attributes of the tabular data sets, except the class attribute.

To implement probabilistic $k$-anonymity on the tabular data sets, we used the MDAV microaggregation algorithm~\cite{domingo2005ordinal} 
to compute clusters of at least $k$ records.
Within each cluster, we applied column-wise permutation: each quasi-identifier attribute column is independently permuted.
For probabilistic $k$-anonymity on the CIFAR10 image data set, we extracted 512-dimensional latent features from an ImageNet-pretrained ResNet18 model and used them to form MDAV clusters of at least $k$  images. 
Within each cluster, we then applied 
pixel-wise permutation, independently shuffling each pixel position across images in the cluster.

For DP with tabular data sets, we used the Laplace mechanism~\cite{dwork2006calibrating} for numerical attributes and the exponential mechanism~\cite{mcsherry2007mechanism} for categorical attributes.
For DP on the Adult data set, which contains several non-ordinal categorical attributes, we leveraged the unsupervised training capabilities of TabNet~\cite{arik2021tabnet} to generate embeddings for these attributes, which allowed us to encode each non-ordinal categorical attribute into a $10$-dimensional embedding vector.
Then, we used the cosine similarity between the embeddings of attributes as a utility function for the exponential mechanism. 
For the CIFAR10 image data set, we enforced DP via the DP-Pix methodology described in~\cite{fan2019differential}. 
DP-Pix first pixelizes the image by averaging pixel values in blocks of size $b \times b$ to reduce sensitivity and then adds noise to the pixels using the Laplace mechanism based on the global sensitivity $255m/b^2$. The parameter $m$ defines the number of different pixels between neighboring images. 
We used $m=16$, as suggested by the DP-Pix author, and $b=4$, which is appropriate for this image resolution~\cite{fan2019differential}.

For the main experiments, we used $k=30$ with probabilistic $k$-anonymity (corresponding to a maximum re-identification probability of $1/k \approx 0.033$, well below the 
threshold of $0.09$ recommended by both the European Medicines Agency 
and Health Canada for clinical data~\cite{ema2025,healthcanada2019}); 
to study the sensitivity to the privacy parameter, we also swept 
$k \in \{3, 5, 10, 20, 30, 50, 80, 100, 200, 300, 500, 1000\}$.
For DP, we used $\epsilon=1$, which is considered a robust value \cite{dwork2011firm}; to study the sensitivity to the privacy parameter, we also swept $\epsilon \in \{0.25, 0.5, 1, 2, 5, 10, 20, 50, 100, 150, 200, 300\}$. 
For all data sets, we enforced DP in its 
pure form, without any $\delta$-based relaxation, {\em i.e.}, $\delta=0$.
Section~\ref{sec:sensitivity} provides further empirical analysis for the impact of \(k\) and \(\epsilon\).

\paragraph{ML model training settings} For each tabular data set, we trained an MLP model and an XGBoost model from scratch on the entire training set $\mathbf{D}$. 
We used cross-entropy loss and the Adam optimizer with a cosine learning rate schedule to train all MLP models. 
The two hidden layers of each MLP were configured with 
128 neurons for the Adult and Heart benchmarks and 256 neurons for the Credit benchmark due to its larger training set size and class imbalance.

For CIFAR10, we trained the ResNet18 model using cross-entropy loss and the SGD optimizer with a cosine learning rate schedule.
The specific hyperparameters used during the training of these benchmarks are detailed in Table~\ref{tab:training_hyperparameters}.

\paragraph{Unlearning method settings} For SISA, we split the training set into 5 disjoint shards (each containing 10 slices) and applied the SISA training procedure using the training hyperparameters presented in Table~\ref{tab:training_hyperparameters}.

For Certified-SP~\cite{koloskova2025certified} unlearning, we applied the gradient-clipping variant
with $\epsilon=1$, $\delta=10^{-5}$ to the original trained models, following the settings of the original paper.
Certified-SP is not applied to XGBoost as it requires gradient-based optimization
during unlearning, which is incompatible with tree-based models.

For our EUPG method with tabular data, we obtained the base private models $M^k$ and $M^{\epsilon}$ by training them on the $\mathbf{D}^k$ and $\mathbf{D}^{\epsilon}$ protected data sets, respectively, using the same training hyperparameters in Table~\ref{tab:training_hyperparameters}. 
For EUPG on CIFAR10, we also trained on $\mathbf{D}^k$ and $\mathbf{D}^{\epsilon}$.
We then fine-tuned $M^k$ and $M^{\epsilon}$ on $\mathbf{D}$ for 5 epochs (MLP and ResNet18) or 5 additional estimators (XGBoost) to obtain $M^k_\mathbf{D}$ and $M^{\epsilon}_\mathbf{D}$, using the learning rates and batch sizes in Table~\ref{tab:training_hyperparameters}.

\begin{table}
\centering
\caption{Training hyperparameters}
\label{tab:training_hyperparameters}
\resizebox{\columnwidth}{!}{%
\begin{tabular}{lcl} 
\toprule
\multicolumn{1}{c}{Data set}    & Model                       & \multicolumn{1}{c}{Hyperparameters}                                                                   \\ 
\midrule
\multirow{2}{*}{Adult income}  & MLP                         & BS:256, LR:1e-2, Epochs:50, WD:1e-4  \\
                               & \multicolumn{1}{l}{XGBoost} & Estimators:100, Depth:7, LR:0.5, $\lambda$:5                                                           \\ 
\midrule
\multirow{2}{*}{Heart disease} & MLP                         & BS:256, LR:1e-2, Epochs:50, WD:1e-5  \\
                               & \multicolumn{1}{l}{XGBoost} & Estimators:100, Depth:7, LR:0.5, $\lambda$:5                                                            \\ 
\midrule
\multirow{2}{*}{Credit~}       & MLP                         & BS:256, LR:1e-2, Epochs:50, WD:1e-5   \\
                               & \multicolumn{1}{l}{XGBoost} & Estimators:100, Depth:7, LR:0.5, $\lambda$:5                                                            \\ 
\midrule
\multirow{2}{*}{CIFAR10~}       & \multirow{2}{*}{ResNet18}  & BS:256, Epochs:100, WD:5e-4\\
                                &                             &  LR:(1e-1 train, 5e-2 fine-tune)    \\

\bottomrule
\end{tabular}}%
\end{table}

\paragraph{Forgetting request settings} We randomly sampled a forget set $\mathbf{D}_f$ from the original training set $\mathbf{D}$ using a given forgetting ratio ({\em e.g.}, 5\%). 
After subtracting $\mathbf{D}_f$ from $\mathbf{D}$, the remaining training points constituted the retain set $\mathbf{D}_r$. 
To forget $\mathbf{D}_f$ with privacy guarantees, we fine-tuned $M^k$ and $M^{\epsilon}$ on $\mathbf{D}_r$ to obtain $M^k_{\mathbf{D}_r}$ and $M^{\epsilon}_{\mathbf{D}_r}$, respectively.
We used the learning rates and batch sizes in Table~\ref{tab:training_hyperparameters}.
For MLP models, we fine-tuned for 5 epochs using the cosine learning rate schedule.
For ResNet18, we fine-tuned for 5 epochs using the OneCycleLR schedule.
For XGBoost, we added 5 estimators and fine-tuned the models with those added estimators.

For all training, unlearning, and attack experiments, the results are reported as mean $\pm$ standard deviation over 5 runs.

\subsection{Main results}
\label{sec:results}

We evaluated all methods on two dimensions: first, \emph{unlearning quality} ---utility
preservation (test accuracy or AUC) and forgetting effectiveness (RMIA AUC and
TPR@1\%FPR)--- and, second, \emph{unlearning efficiency} ---unlearning time relative to full
retraining.

\paragraph{Unlearning quality.}
Tables~\ref{tab:quality_mlp} and~\ref{tab:quality_xgboost} report the results for MLP and XGBoost on Adult, Heart, and Credit at a forget ratio of~$5\%$.
For a fair and direct comparison with Certified-SP~\cite{koloskova2025certified}, we report EUPG results at the same DP budget ({\em i.e.}, $\epsilon=1$), which is considered `safe' as discussed in~\cite{dwork2011firm}.

For MLP, all baselines (retraining, SISA, and Certified-SP) and both EUPG variants
(with prob. $k$-anonymity and DP) achieved similar utility both before and after unlearning.
Across all tabular data sets, all methods yielded post-unlearning MIA AUC values close to random guessing, indicating effective removal of membership signal for the forget set.
By contrast, Certified-SP was only applicable to neural networks (Table~\ref{tab:quality_xgboost}).
Before unlearning, the MIA results of the EUPG variants were substantially lower than those of the corresponding full-data model and SISA.
This suggests that fine-tuning the protected base model on the full original data set (\(\mathbf{D}\)) for a few epochs ({\em e.g.}, 5) can recover utility quickly without substantially reintroducing 
membership signal.

\begin{table*}[t!]
\centering
\small
\caption{Utility and MIA resistance before/after unlearning (MLP, forget ratio~$= 5\%$, 5 fine-tuning epochs, mean~$\pm$~std over 5 runs). Utility is accuracy~(\%) for Adult and Heart, AUC~(\%) for Credit. In italics, the retrain baseline. 
    Among the remaining methods, boldface denotes the best and underlining denotes the second best for each metric.}
\label{tab:quality_mlp}
\resizebox{\textwidth}{!}{%
\begin{tabular}{l l c c c c c c}
\toprule
\multirow{2}{*}{Data set} & \multirow{2}{*}{Method} & \multicolumn{2}{c}{Utility\,$\uparrow$} & \multicolumn{2}{c}{MIA~AUC\,$\downarrow$} & \multicolumn{2}{c}{TPR@1\%\,$\downarrow$} \\
\cmidrule(lr){3-4}\cmidrule(lr){5-6}\cmidrule(lr){7-8}
 &  & Before & After & Before & After & Before & After \\
\midrule
\multirow{5}{*}{Adult}
 & \emph{Retrain}
 & $\mathit{85.35}\,{\scriptscriptstyle\pm}\,\mathit{0.08}$
 & $\mathit{85.21}\,{\scriptscriptstyle\pm}\,\mathit{0.10}$
 & $\mathit{55.49}\,{\scriptscriptstyle\pm}\,\mathit{0.91}$
 & $\mathit{49.66}\,{\scriptscriptstyle\pm}\,\mathit{0.97}$
 & $\mathit{3.78}\,{\scriptscriptstyle\pm}\,\mathit{1.24}$
 & $\mathit{0.86}\,{\scriptscriptstyle\pm}\,\mathit{0.32}$ \\
 & SISA
 & $85.16\,{\scriptscriptstyle\pm}\,0.07$
 & $85.29\,{\scriptscriptstyle\pm}\,0.08$
 & $55.86\,{\scriptscriptstyle\pm}\,1.06$
 & $\mathbf{49.75}\,{\scriptscriptstyle\pm}\,\mathbf{0.66}$
 & $2.61\,{\scriptscriptstyle\pm}\,0.47$
 & $\mathbf{0.93}\,{\scriptscriptstyle\pm}\,\mathbf{0.41}$ \\
 & Certified-SP ($\epsilon$=1, $\delta$=1e-5)
 & $\underline{85.35\,{\scriptscriptstyle\pm}\,0.08}$
 & $\mathbf{85.63}\,{\scriptscriptstyle\pm}\,\mathbf{0.04}$
 & $55.49\,{\scriptscriptstyle\pm}\,0.91$
 & $50.01\,{\scriptscriptstyle\pm}\,0.89$
 & $3.78\,{\scriptscriptstyle\pm}\,1.24$
 & $1.15\,{\scriptscriptstyle\pm}\,0.25$ \\
 & EUPG ($k$=30)
 & $\mathbf{85.64}\,{\scriptscriptstyle\pm}\,\mathbf{0.08}$
 & $85.58\,{\scriptscriptstyle\pm}\,0.10$
 & $\underline{51.48\,{\scriptscriptstyle\pm}\,0.97}$
 & $\underline{49.79\,{\scriptscriptstyle\pm}\,0.78}$
 & $\underline{1.54\,{\scriptscriptstyle\pm}\,0.22}$
 & $1.15\,{\scriptscriptstyle\pm}\,0.21$ \\
 & EUPG ($\epsilon$=1, $\delta$=0)
 & $\mathbf{85.64}\,{\scriptscriptstyle\pm}\,\mathbf{0.14}$
 & $\underline{85.60\,{\scriptscriptstyle\pm}\,0.16}$
 & $\mathbf{51.35}\,{\scriptscriptstyle\pm}\,\mathbf{0.60}$
 & $49.91\,{\scriptscriptstyle\pm}\,0.83$
 & $\mathbf{1.43}\,{\scriptscriptstyle\pm}\,\mathbf{0.14}$
 & $\underline{0.97\,{\scriptscriptstyle\pm}\,0.21}$ \\

\midrule
\multirow{5}{*}{Heart}
 & \emph{Retrain}
 & $\mathit{73.38}\,{\scriptscriptstyle\pm}\,\mathit{0.04}$
 & $\mathit{73.29}\,{\scriptscriptstyle\pm}\,\mathit{0.17}$
 & $\mathit{53.81}\,{\scriptscriptstyle\pm}\,\mathit{0.72}$
 & $\mathit{49.96}\,{\scriptscriptstyle\pm}\,\mathit{0.50}$
 & $\mathit{2.54}\,{\scriptscriptstyle\pm}\,\mathit{0.30}$
 & $\mathit{1.05}\,{\scriptscriptstyle\pm}\,\mathit{0.29}$ \\
 & SISA
 & $72.98\,{\scriptscriptstyle\pm}\,0.07$
 & $72.99\,{\scriptscriptstyle\pm}\,0.27$
 & $56.04\,{\scriptscriptstyle\pm}\,0.44$
 & $\mathbf{50.05}\,{\scriptscriptstyle\pm}\,\mathbf{0.58}$
 & $2.73\,{\scriptscriptstyle\pm}\,0.56$
 & $1.18\,{\scriptscriptstyle\pm}\,0.26$ \\
 & Certified-SP ($\epsilon$=1, $\delta$=1e-5)
 & $\mathbf{73.38}\,{\scriptscriptstyle\pm}\,\mathbf{0.04}$
 & $\underline{73.33\,{\scriptscriptstyle\pm}\,0.03}$
 & $53.81\,{\scriptscriptstyle\pm}\,0.72$
 & $50.28\,{\scriptscriptstyle\pm}\,0.25$
 & $2.54\,{\scriptscriptstyle\pm}\,0.30$
 & $1.03\,{\scriptscriptstyle\pm}\,0.04$ \\
 & EUPG ($k$=30)
 & $73.25\,{\scriptscriptstyle\pm}\,0.07$
 & $73.29\,{\scriptscriptstyle\pm}\,0.10$
 & $\underline{51.37\,{\scriptscriptstyle\pm}\,0.33}$
 & $\underline{50.23\,{\scriptscriptstyle\pm}\,0.23}$
 & $\underline{1.40\,{\scriptscriptstyle\pm}\,0.15}$
 & $\mathbf{0.99}\,{\scriptscriptstyle\pm}\,\mathbf{0.13}$ \\
 & EUPG ($\epsilon$=1, $\delta$=0)
 & $\underline{73.35\,{\scriptscriptstyle\pm}\,0.10}$
 & $\mathbf{73.37}\,{\scriptscriptstyle\pm}\,\mathbf{0.10}$
 & $\mathbf{50.77}\,{\scriptscriptstyle\pm}\,\mathbf{0.28}$
 & $\underline{50.23\,{\scriptscriptstyle\pm}\,0.35}$
 & $\mathbf{1.15}\,{\scriptscriptstyle\pm}\,\mathbf{0.13}$
 & $\underline{1.00\,{\scriptscriptstyle\pm}\,0.11}$ \\

\midrule
\multirow{5}{*}{Credit}
 & \emph{Retrain}
 & $\mathit{81.46}\,{\scriptscriptstyle\pm}\,\mathit{0.05}$
 & $\mathit{81.44}\,{\scriptscriptstyle\pm}\,\mathit{0.04}$
 & $\mathit{50.43}\,{\scriptscriptstyle\pm}\,\mathit{0.27}$
 & $\mathit{50.02}\,{\scriptscriptstyle\pm}\,\mathit{0.16}$
 & $\mathit{1.47}\,{\scriptscriptstyle\pm}\,\mathit{0.27}$
 & $\mathit{1.01}\,{\scriptscriptstyle\pm}\,\mathit{0.09}$ \\
 & SISA
 & $80.92\,{\scriptscriptstyle\pm}\,0.07$
 & $81.19\,{\scriptscriptstyle\pm}\,0.03$
 & $52.05\,{\scriptscriptstyle\pm}\,0.25$
 & $50.11\,{\scriptscriptstyle\pm}\,0.21$
 & $2.06\,{\scriptscriptstyle\pm}\,0.28$
 & $1.05\,{\scriptscriptstyle\pm}\,0.14$ \\
 & Certified-SP ($\epsilon$=1, $\delta$=1e-5)
 & $\mathbf{81.46}\,{\scriptscriptstyle\pm}\,\mathbf{0.05}$
 & $\mathbf{81.34}\,{\scriptscriptstyle\pm}\,\mathbf{0.06}$
 & $50.43\,{\scriptscriptstyle\pm}\,0.27$
 & $\mathbf{49.98}\,{\scriptscriptstyle\pm}\,\mathbf{0.22}$
 & $1.47\,{\scriptscriptstyle\pm}\,0.27$
 & $\mathbf{0.83}\,{\scriptscriptstyle\pm}\,\mathbf{0.13}$ \\
 & EUPG ($k$=30)
 & $81.26\,{\scriptscriptstyle\pm}\,0.05$
 & $81.26\,{\scriptscriptstyle\pm}\,0.01$
 & $\mathbf{50.33}\,{\scriptscriptstyle\pm}\,\mathbf{0.21}$
 & $\underline{50.04\,{\scriptscriptstyle\pm}\,0.11}$
 & $\underline{1.20\,{\scriptscriptstyle\pm}\,0.16}$
 & $0.99\,{\scriptscriptstyle\pm}\,0.10$ \\
 & EUPG ($\epsilon$=1, $\delta$=0)
 & $\underline{81.37\,{\scriptscriptstyle\pm}\,0.04}$
 & $\underline{81.31\,{\scriptscriptstyle\pm}\,0.02}$
 & $\underline{50.39\,{\scriptscriptstyle\pm}\,0.08}$
 & $50.11\,{\scriptscriptstyle\pm}\,0.14$
 & $\mathbf{1.15}\,{\scriptscriptstyle\pm}\,\mathbf{0.10}$
 & $\underline{0.92\,{\scriptscriptstyle\pm}\,0.09}$ \\
\bottomrule
\end{tabular}}
\end{table*}

%JOSEP2. Paragraph a bit rewritten.
For XGBoost (Table~\ref{tab:quality_xgboost}), EUPG(\(k=30\)) suffered only a slight drop in utility from before to after unlearning relative to Retrain and SISA, while EUPG(\(\epsilon=1\)) showed a somewhat larger drop, especially on the Heart data set.
All methods achieved similarly strong resistance to MIA after unlearning.
At the same time, both EUPG variants were substantially more resistant to MIA than Retrain and SISA before unlearning, with markedly lower MIA AUC and TPR@1\%FPR across all data sets.
After unlearning, both variants achieved MIA AUC values close to \(50\%\) and low TPR@1\%FPR, indicating strong empirical forgetting.
With EUPG($\epsilon=1$), the utility of the model decreased slightly compared to the other methods.
This pattern can be attributed to the sequential structure of the boosted trees.
Because each tree is fitted to the residuals determined by the preceding trees, initial training on protected data can affect the split structure learned throughout the ensemble.
Moreover, since the post-unlearning fine-tuning stage updates only leaf values while keeping the tree structure fixed, suboptimal splits introduced during protected pre-training cannot be corrected afterward, which may slightly limit utility recovery.
Even so, the gain in empirical forgetting relative to pre-unlearning is substantial.

Overall, these results show that EUPG provides a good utility--privacy trade-off for XGBoost, with strong resistance to MIA already before unlearning and strong empirical forgetting afterwards.

\begin{table*}[t!]
\centering
\small
\caption{Utility and MIA resistance before/after unlearning (XGBoost, forget ratio~$= 5\%$, 5 fine-tuning epochs, mean~$\pm$~std over 5 runs). Utility is accuracy~(\%) for Adult and Heart, AUC~(\%) for Credit. In italics, the retrain baseline. Among the remaining methods, boldface denotes the best and underlining denotes the second best for each metric.}
\label{tab:quality_xgboost}
\resizebox{\textwidth}{!}{%
\begin{tabular}{l l c c c c c c}
\toprule
\multirow{2}{*}{Data set} & \multirow{2}{*}{Method} & \multicolumn{2}{c}{Utility\,$\uparrow$} & \multicolumn{2}{c}{MIA~AUC\,$\downarrow$} & \multicolumn{2}{c}{TPR@1\%\,$\downarrow$} \\
\cmidrule(lr){3-4}\cmidrule(lr){5-6}\cmidrule(lr){7-8}
 &  & Before & After & Before & After & Before & After \\
\midrule
\multirow{4}{*}{Adult}
 & \emph{Retrain}
 & $\mathit{86.88}\,{\scriptscriptstyle\pm}\,\mathit{0.04}$
 & $\mathit{87.00}\,{\scriptscriptstyle\pm}\,\mathit{0.08}$
 & $\mathit{57.92}\,{\scriptscriptstyle\pm}\,\mathit{0.75}$
 & $\mathit{49.44}\,{\scriptscriptstyle\pm}\,\mathit{0.72}$
 & $\mathit{4.92}\,{\scriptscriptstyle\pm}\,\mathit{1.12}$
 & $\mathit{0.89}\,{\scriptscriptstyle\pm}\,\mathit{0.47}$ \\
 & SISA
 & $\mathbf{87.01}\,{\scriptscriptstyle\pm}\,\mathbf{0.05}$
 & $\mathbf{87.03}\,{\scriptscriptstyle\pm}\,\mathbf{0.07}$
 & $54.38\,{\scriptscriptstyle\pm}\,0.45$
 & $\mathbf{49.67}\,{\scriptscriptstyle\pm}\,\mathbf{0.66}$
 & $2.43\,{\scriptscriptstyle\pm}\,0.54$
 & $\mathbf{0.83}\,{\scriptscriptstyle\pm}\,\mathbf{0.30}$ \\
 & EUPG ($k$=30)
 & $\underline{86.44\,{\scriptscriptstyle\pm}\,0.12}$
 & $\underline{86.45\,{\scriptscriptstyle\pm}\,0.17}$
 & $\mathbf{50.72}\,{\scriptscriptstyle\pm}\,\mathbf{0.89}$
 & $50.42\,{\scriptscriptstyle\pm}\,0.84$
 & $\underline{1.31\,{\scriptscriptstyle\pm}\,0.24}$
 & $1.21\,{\scriptscriptstyle\pm}\,0.31$ \\
 & EUPG ($\epsilon$=1, $\delta$=0)
 & $85.37\,{\scriptscriptstyle\pm}\,0.23$
 & $85.39\,{\scriptscriptstyle\pm}\,0.22$
 & $\underline{50.83\,{\scriptscriptstyle\pm}\,0.44}$
 & $\underline{50.20\,{\scriptscriptstyle\pm}\,0.52}$
 & $\mathbf{1.14}\,{\scriptscriptstyle\pm}\,\mathbf{0.26}$
 & $\underline{1.06\,{\scriptscriptstyle\pm}\,0.20}$ \\
\midrule
\multirow{4}{*}{Heart}
 & \emph{Retrain}
 & $\mathit{72.43}\,{\scriptscriptstyle\pm}\,\mathit{0.17}$
 & $\mathit{72.45}\,{\scriptscriptstyle\pm}\,\mathit{0.25}$
 & $\mathit{64.58}\,{\scriptscriptstyle\pm}\,\mathit{0.74}$
 & $\mathit{50.11}\,{\scriptscriptstyle\pm}\,\mathit{0.55}$
 & $\mathit{6.10}\,{\scriptscriptstyle\pm}\,\mathit{0.43}$
 & $\mathit{1.03}\,{\scriptscriptstyle\pm}\,\mathit{0.14}$ \\
 & SISA
 & $\mathbf{73.06}\,{\scriptscriptstyle\pm}\,\mathbf{0.13}$
 & $\mathbf{73.09}\,{\scriptscriptstyle\pm}\,\mathbf{0.08}$
 & $60.28\,{\scriptscriptstyle\pm}\,0.40$
 & $\underline{50.22\,{\scriptscriptstyle\pm}\,0.41}$
 & $3.24\,{\scriptscriptstyle\pm}\,0.23$
 & $\mathbf{0.97}\,{\scriptscriptstyle\pm}\,\mathbf{0.15}$ \\
 & EUPG ($k$=30)
 & $\underline{72.30\,{\scriptscriptstyle\pm}\,0.18}$
 & $\underline{72.21\,{\scriptscriptstyle\pm}\,0.29}$
 & $\underline{51.51\,{\scriptscriptstyle\pm}\,0.47}$
 & $50.51\,{\scriptscriptstyle\pm}\,0.40$
 & $\underline{1.38\,{\scriptscriptstyle\pm}\,0.32}$
 & $1.07\,{\scriptscriptstyle\pm}\,0.20$ \\
 & EUPG ($\epsilon$=1, $\delta$=0)
 & $70.72\,{\scriptscriptstyle\pm}\,0.18$
 & $70.74\,{\scriptscriptstyle\pm}\,0.06$
 & $\mathbf{51.36}\,{\scriptscriptstyle\pm}\,\mathbf{0.50}$
 & $\mathbf{50.12}\,{\scriptscriptstyle\pm}\,\mathbf{0.60}$
 & $\mathbf{1.09}\,{\scriptscriptstyle\pm}\,\mathbf{0.06}$
 & $\underline{1.04\,{\scriptscriptstyle\pm}\,0.08}$ \\
\midrule
\multirow{4}{*}{Credit}
 & \emph{Retrain}
 & $\mathit{82.90}\,{\scriptscriptstyle\pm}\,\mathit{0.18}$
 & $\mathit{82.65}\,{\scriptscriptstyle\pm}\,\mathit{0.21}$
 & $\mathit{56.83}\,{\scriptscriptstyle\pm}\,\mathit{0.51}$
 & $\mathit{49.63}\,{\scriptscriptstyle\pm}\,\mathit{0.48}$
 & $\mathit{4.94}\,{\scriptscriptstyle\pm}\,\mathit{0.47}$
 & $\mathit{1.08}\,{\scriptscriptstyle\pm}\,\mathit{0.17}$ \\
 & SISA
 & $\mathbf{83.12}\,{\scriptscriptstyle\pm}\,\mathbf{0.12}$
 & $\mathbf{83.03}\,{\scriptscriptstyle\pm}\,\mathbf{0.13}$
 & $55.02\,{\scriptscriptstyle\pm}\,0.39$
 & $\mathbf{49.97}\,{\scriptscriptstyle\pm}\,\mathbf{0.24}$
 & $3.98\,{\scriptscriptstyle\pm}\,0.39$
 & $\mathbf{0.98}\,{\scriptscriptstyle\pm}\,\mathbf{0.28}$ \\
 & EUPG ($k$=30)
 & $\underline{82.16\,{\scriptscriptstyle\pm}\,0.20}$
 & $\underline{82.18\,{\scriptscriptstyle\pm}\,0.15}$
 & $\mathbf{50.59}\,{\scriptscriptstyle\pm}\,\mathbf{0.35}$
 & $\underline{49.98\,{\scriptscriptstyle\pm}\,0.56}$
 & $\underline{1.26\,{\scriptscriptstyle\pm}\,0.21}$
 & $1.10\,{\scriptscriptstyle\pm}\,0.16$ \\
 & EUPG ($\epsilon$=1, $\delta$=0)
 & $80.93\,{\scriptscriptstyle\pm}\,2.57$
 & $81.14\,{\scriptscriptstyle\pm}\,2.09$
 & $\underline{51.13\,{\scriptscriptstyle\pm}\,0.21}$
 & $50.17\,{\scriptscriptstyle\pm}\,0.19$
 & $\mathbf{1.20}\,{\scriptscriptstyle\pm}\,\mathbf{0.19}$
 & $\underline{1.06\,{\scriptscriptstyle\pm}\,0.13}$ \\
\bottomrule
\end{tabular}}
\end{table*}

Table~\ref{tab:quality_resnet18} reports the results for ResNet18 on CIFAR10 at a forget ratio of~$5\%$.
Again, all methods achieved similarly strong resistance to MIA after unlearning.
Before unlearning, both variants of EUPG were superior in MIA resistance.
Baseline retrain achieved $94.25\%$ test accuracy while SISA (\(S=5\)) suffered a significant loss in utility, 
with accuracy after unlearning falling to $69.19\%$.
This probably reflects a combination of limited per-shard data, fragmentation across slices, 
and the inability of the ensemble to compensate for weak sub-models.
Certified-SP~\cite{koloskova2025certified} performed even worse, 
collapsing to $45.91\%$ test accuracy after unlearning at \(\epsilon=1\).
Koloskova {\em et al.}~\cite{koloskova2025certified} note that the effectiveness of their method is constrained by the curse of dimensionality inherent in applying differential privacy in the model space.
In this setting, the combination of clipping/noise in a full ResNet18 and a small post-unlearning fine-tuning budget appears insufficient to recover a high-quality representation.
This highlights that our \emph{data pre-protection} approach can be a more practical route to unlearning in high-dimensional models: EUPG(\(k=30\)) achieved the best post-unlearning utility at $85.40\%$, 
followed by EUPG(\(\epsilon=1\)) at $83.17\%$.

Overall, EUPG(\(k=30\)) offered the best utility--privacy trade-off on CIFAR10 before and after unlearning.

\begin{table*}[t!]
\centering
\small
\caption{Utility and MIA resistance before/after unlearning (CIFAR10-ResNet18, forget ratio~$= 5\%$, 5 fine-tuning epochs, mean~$\pm$~std over 5 runs). In italics, the retrain baseline. 
    Among the remaining methods, boldface denotes the best and underlining denotes the second best for each metric.}
\label{tab:quality_resnet18}
\resizebox{\textwidth}{!}{%
\begin{tabular}{l l c c c c c c}
\toprule
\multirow{2}{*}{Data set} & \multirow{2}{*}{Method} & \multicolumn{2}{c}{Utility\,$\uparrow$} & \multicolumn{2}{c}{MIA~AUC\,$\downarrow$} & \multicolumn{2}{c}{TPR@1\%\,$\downarrow$} \\
\cmidrule(lr){3-4}\cmidrule(lr){5-6}\cmidrule(lr){7-8}
 &  & Bef. & Aft. & Bef. & Aft. & Bef. & Aft. \\
\midrule
\multirow{5}{*}{CIFAR10}
 & \emph{Retrain}
 & $\mathit{94.70}\,{\scriptscriptstyle\pm}\,\mathit{0.19}$
 & $\mathit{94.25}\,{\scriptscriptstyle\pm}\,\mathit{0.23}$
 & $\mathit{57.71}\,{\scriptscriptstyle\pm}\,\mathit{0.43}$
 & $\mathit{50.04}\,{\scriptscriptstyle\pm}\,\mathit{0.28}$
 & $\mathit{4.18}\,{\scriptscriptstyle\pm}\,\mathit{0.36}$
 & $\mathit{0.82}\,{\scriptscriptstyle\pm}\,\mathit{0.09}$ \\
 & SISA
 & $\underline{87.34\,{\scriptscriptstyle\pm}\,0.23}$
 & $69.19\,{\scriptscriptstyle\pm}\,0.21$
 & $52.05\,{\scriptscriptstyle\pm}\,0.38$
 & $48.07\,{\scriptscriptstyle\pm}\,0.82$
 & $1.63\,{\scriptscriptstyle\pm}\,0.31$
 & $\mathbf{0.79}\,{\scriptscriptstyle\pm}\,\mathbf{0.24}$ \\
 & Certified-SP ($\epsilon$=1, $\delta$=1e-5)
 & $\mathbf{94.70}\,{\scriptscriptstyle\pm}\,\mathbf{0.19}$
 & $45.91\,{\scriptscriptstyle\pm}\,0.65$
 & $57.71\,{\scriptscriptstyle\pm}\,0.43$
 & $48.47\,{\scriptscriptstyle\pm}\,0.36$
 & $4.18\,{\scriptscriptstyle\pm}\,0.36$
 & $1.02\,{\scriptscriptstyle\pm}\,0.11$ \\
 & EUPG ($k$=30)
 & $85.89\,{\scriptscriptstyle\pm}\,0.18$
 & $\mathbf{85.40}\,{\scriptscriptstyle\pm}\,\mathbf{0.27}$
 & $\underline{49.64\,{\scriptscriptstyle\pm}\,0.55}$
 & $\mathbf{47.41}\,{\scriptscriptstyle\pm}\,\mathbf{0.43}$
 & $\underline{1.25\,{\scriptscriptstyle\pm}\,0.29}$
 & $\underline{1.00\,{\scriptscriptstyle\pm}\,0.18}$ \\
 & EUPG ($\epsilon$=1, $\delta$=0)
 & $83.54\,{\scriptscriptstyle\pm}\,0.35$
 & $\underline{83.17\,{\scriptscriptstyle\pm}\,0.24}$
 & $\mathbf{49.18}\,{\scriptscriptstyle\pm}\,\mathbf{0.31}$
 & $\underline{47.52\,{\scriptscriptstyle\pm}\,0.57}$
 & $\mathbf{1.15}\,{\scriptscriptstyle\pm}\,\mathbf{0.16}$
 & $1.03\,{\scriptscriptstyle\pm}\,0.20$ \\
\bottomrule
\end{tabular}}
\end{table*}

\paragraph{Unlearning efficiency.}
Table~\ref{tab:efficiency_headtohead} compares the unlearning time across methods and data sets.
EUPG was the most efficient family of methods overall: both variants were several times faster than Retrain across all settings,
outperforming SISA by a large margin, and almost as fast as Certified-SP in every neural-network case.

On neural networks, EUPG achieved average speedups of \(14.76\times\) (EUPG\(_k\))
and \(14.84\times\) (EUPG\(_\epsilon\)) versus \(14.04\times\) for Certified-SP.
On XGBoost, speedups of \(2.74\times\) and \(2.47\times\) were achieved with absolute
unlearning times of \(0.030\)--\(0.048\)\,s; the smaller relative gains reflect the
already very low retraining baseline.
Critically, EUPG was the only method that combined this efficiency with applicability across all
ML model types, whereas Certified-SP is restricted to neural
networks and showed severe utility degradation on CIFAR10
(Table~\ref{tab:quality_resnet18}).

SISA was slower than full retraining in both model families (\(0.62\times\) and
\(0.33\times\)) because the randomly sampled forget set forced all five shards to retrain
from slice~0. SISA can only get reasonable runtimes  when forget sets are small and concentrated in
recent slices.

\begin{table*}[t!]
\centering
\small
\caption{Unlearning runtime (seconds, forget ratio~$= 5\%$, mean~$\pm$~std over 5 runs). Speedup is relative to Retrain. 
In italics, the retrain baseline. Among the remaining methods, boldface denotes the best method and underlining the second best per column. NA denotes not applicable.}
\label{tab:efficiency_headtohead}
\setlength{\tabcolsep}{4pt}
\resizebox{\textwidth}{!}{%
\begin{tabular}{l cccc c cccc}
\toprule
 & \multicolumn{5}{c}{Neural Network} & \multicolumn{4}{c}{XGBoost} \\
\cmidrule(lr){2-6}\cmidrule(lr){7-10}
Method & Adult & Heart & Credit & CIFAR10 & Speedup & Adult & Heart & Credit & Speedup \\
\midrule
\emph{Retrain}
& $\mathit{17.99}\,{\scriptscriptstyle\pm}\,\mathit{0.13}$
& $\mathit{31.75}\,{\scriptscriptstyle\pm}\,\mathit{1.92}$
& $\mathit{52.54}\,{\scriptscriptstyle\pm}\,\mathit{1.37}$
& $\mathit{417.32}\,{\scriptscriptstyle\pm}\,\mathit{0.44}$
& $\mathit{1.00\times}$
& $\mathit{0.096}\,{\scriptscriptstyle\pm}\,\mathit{0.023}$
& $\mathit{0.088}\,{\scriptscriptstyle\pm}\,\mathit{0.001}$
& $\mathit{0.113}\,{\scriptscriptstyle\pm}\,\mathit{0.003}$
& $\mathit{1.00\times}$ \\
\addlinespace
SISA
& $37.96\,{\scriptscriptstyle\pm}\,0.96$
& $52.57\,{\scriptscriptstyle\pm}\,6.06$
& $82.88\,{\scriptscriptstyle\pm}\,4.59$
& $530.35\,{\scriptscriptstyle\pm}\,1.95$
& $0.62\times$
& $0.216\,{\scriptscriptstyle\pm}\,0.006$
& $0.228\,{\scriptscriptstyle\pm}\,0.005$
& $0.661\,{\scriptscriptstyle\pm}\,0.007$
& $0.33\times$ \\
\addlinespace
Certified-SP ($\epsilon$=1, $\delta$=1e-5)
& $\underline{1.50\,{\scriptscriptstyle\pm}\,0.05}$
& $1.84\,{\scriptscriptstyle\pm}\,0.08$
& $\underline{5.59\,{\scriptscriptstyle\pm}\,0.25}$
& $23.81\,{\scriptscriptstyle\pm}\,0.18$
& $14.04\times$
& NA & NA & NA & NA \\
\addlinespace
EUPG ($k$=30)
& $\mathbf{1.24}\,{\scriptscriptstyle\pm}\,\mathbf{0.11}$
& $\underline{1.82\,{\scriptscriptstyle\pm}\,0.08}$
& $\underline{5.59\,{\scriptscriptstyle\pm}\,0.19}$
& $\underline{23.62\,{\scriptscriptstyle\pm}\,0.55}$
& $\underline{14.76\times}$
& $\mathbf{0.030}\,{\scriptscriptstyle\pm}\,\mathbf{0.002}$
& $\mathbf{0.033}\,{\scriptscriptstyle\pm}\,\mathbf{0.003}$
& $\underline{0.048\,{\scriptscriptstyle\pm}\,0.002}$
& $\mathbf{2.74\times}$ \\
\addlinespace
EUPG ($\epsilon$=1, $\delta$=0)
& $\mathbf{1.24}\,{\scriptscriptstyle\pm}\,\mathbf{0.10}$
& $\mathbf{1.80}\,{\scriptscriptstyle\pm}\,\mathbf{0.08}$
& $\mathbf{5.55}\,{\scriptscriptstyle\pm}\,\mathbf{0.25}$
& $\mathbf{23.53}\,{\scriptscriptstyle\pm}\,\mathbf{0.09}$
& $\mathbf{14.84\times}$
& $\underline{0.032\,{\scriptscriptstyle\pm}\,0.002}$
& $\underline{0.045\,{\scriptscriptstyle\pm}\,0.002}$
& $\mathbf{0.046}\,{\scriptscriptstyle\pm}\,\mathbf{0.001}$
& $\underline{2.46\times}$ \\
\addlinespace
\bottomrule
\end{tabular}}
\end{table*}

\paragraph{Storage and inference overhead.}
SISA requires storing $S \times R$ checkpoints (50 in our setup) and incurs an inference
overhead of $(S-1)T$ per query, degrading latency as $S$ grows.
EUPG and Certified-SP, on the other hand, each maintain exactly two model versions: EUPG retains the
protected base model alongside its fine-tuned counterpart, while Certified-SP retains
the pre-unlearning checkpoint as the fixed starting point for each noisy fine-tuning
pass.
Neither adds inference overhead beyond the original model's forward pass.

\paragraph{Summary.}
Overall, EUPG achieved post-unlearning MIA resistance comparable to
exact and certified unlearning baselines across all tabular settings, while uniquely
providing a strong empirical MIA resistance \emph{before}
any unlearning request is made.
On CIFAR10, EUPG is the only method that preserved both utility and forgetting
effectiveness, with SISA and Certified-SP suffering from severe utility degradation for the
same fine-tuning budget.
At the same time, EUPG was consistently the fastest method across all model
families and data sets, achieving a speedup up to $14.84\times$ over full retraining while
requiring only two stored model versions and no inference overhead.

\subsection{Sensitivity analysis of hyperparameters}
\label{sec:sensitivity}

We next study how the anonymity parameter $k$ of probabilistic $k$-anonymity, the DP budget $\epsilon$, the forget ratio,
and the post-fine-tuning epoch count affect the utility--forgetting--efficiency trade-off.
The $k$ and $\epsilon$ sweeps use Heart with XGBoost, which exhibits the sharpest
privacy--utility trade-offs among our tabular benchmarks
(cf.\ Table~\ref{tab:quality_xgboost}), providing a conservative view of EUPG's
sensitivity.

\begin{figure}[t]
\centering
\subfloat[$k$-anonymous data]{
    \includegraphics[width=0.47\columnwidth]{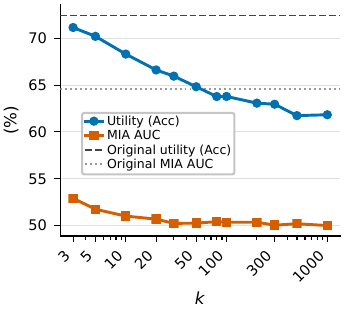}
    \label{fig:k_impact}}
\hfill
\subfloat[DP data]{
    \includegraphics[width=0.47\columnwidth]{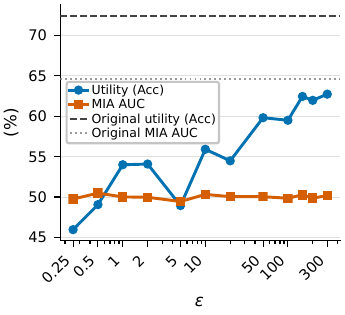}
    \label{fig:e_impact_heart}}

\caption{Impact of privacy parameters ($k$ and $\epsilon$) on utility and 
forgetting for the Heart data set with XGBoost, using the protected base 
model before fine-tuning on original data.}
\label{fig:privacy_params_heart}
\end{figure}

\paragraph{Impact of $k$.}
\label{sec:k_impact}
Figure~\ref{fig:k_impact} shows the impact of $k$ on utility and MIA AUC for the
XGBoost model on the probabilistically $k$-anonymous Heart data set, before fine-tuning.
Utility decreased monotonically from \(\approx71\%\) at \(k=3\) to \(\approx62\%\) at
\(k=1000\).
MIA AUC, on the other hand, dropped sharply from \(\approx65\%\) (dotted line, original data MIA) to \(\approx53\%\) at
\(k=3\), then gradually to \(\approx50\%\) by \(k=30\), after which it largely
saturated: larger \(k\) values continue to reduce utility with almost no additional
privacy benefit, because MIA accuracy already reached near-random guess figures.
This identifies \(k=30\) as the practical operating point, which we used for all experiments.

% \begin{figure}
% \centering
% \setkeys{Gin}{width=\linewidth}
% \begin{subfigure}{0.4\textwidth}
% \includegraphics{figures/k_sweep_heart_xgboost.pdf}
% \caption{Impact of $k$ for the XGBoost model trained on the probabilistically
% $k$-anonymous Heart data set}
% \label{fig:k_impact}
% \end{subfigure}%
% \hfil
% \begin{subfigure}{0.4\textwidth}
% \includegraphics{figures/eps_sweep_heart_xgboost.pdf}
% \caption{Impact of $\epsilon$ for the XGBoost model trained on the differentially
% private Heart data set}
% \label{fig:e_impact_heart}
% \end{subfigure}%
% \caption{Impact of privacy parameters on utility and forgetting for Heart data set, XGBoost and
% protected base model before fine-tuning on original data}
% \end{figure}

\paragraph{Impact of $\epsilon$.}
\label{sec:eps_impact}
Figure~\ref{fig:e_impact_heart} shows the impact of \(\epsilon\) on utility and MIA
AUC for the same XGBoost model on the Heart data set, before fine-tuning.
Utility showed an overall upward trend from \(\approx46\%\) at \(\epsilon=0.25\) to
\(\approx63\%\) at \(\epsilon=300\), though non-monotonically: a visible dip around
\(\epsilon \in [2,10]\) likely reflects uneven budget allocation across numerical and
categorical attributes under sequential composition at intermediate budgets.
In contrast, MIA AUC remained consistently close to \(50\%\) across the entire range,
including at very large \(\epsilon\) where perturbation is negligible.
This indicates that even mild perturbation suffices to suppress empirical membership
signal, and that \(\epsilon=1\) is chosen primarily on utility grounds rather than
empirical privacy necessity ---while still offering a meaningful formal guarantee
consistent with the `safe' regime of~\cite{dwork2011firm}.

\paragraph{Impact of fine-tuning epochs.}
Table~\ref{tab:ft_epochs} shows how post-unlearning fine-tuning epochs affected utility
and MIA resistance on CIFAR10-ResNet18 at \(\epsilon=1\).
EUPG recovered utility substantially faster than Certified-SP: at 5~epochs EUPG
reached \(83.14\%\) versus \(46.08\%\) for Certified-SP, and at 50~epochs \(92.61\%\)
versus \(70.92\%\), the latter remaining well below the \(94.25\%\) retraining
baseline throughout.
MIA AUC stayed close to random guessing for both methods at all epoch counts.

\begin{table}[t!]
\centering
\small
\caption{Impact of fine-tuning epochs on utility and MIA resistance: ResNet18 / CIFAR10, $\epsilon$=1, forget ratio $= 5\%$, mean~$\pm$~std over 5 runs.
In italics, the retrain baseline. Among the remaining methods, boldface denotes  the best.}
\label{tab:ft_epochs}
\setlength{\tabcolsep}{5pt}
\resizebox{\columnwidth}{!}{%
\begin{tabular}{c cc cc}
\toprule
& \multicolumn{2}{c}{\textbf{EUPG ($\epsilon$=1, $\delta$=0)}} 
& \multicolumn{2}{c}{\textbf{Certified-SP} ($\epsilon$=1, $\delta$=1e-5)} \\
\cmidrule(lr){2-3}\cmidrule(lr){4-5}
\textbf{FT epochs}
  & Utility\,$\uparrow$ & MIA AUC\,$\downarrow$
  & Utility\,$\uparrow$ & MIA AUC\,$\downarrow$ \\
\midrule
\emph{Retrain}
  & \multicolumn{2}{c}{$\mathit{94.25}\,{\scriptscriptstyle\pm}\,\mathit{0.23}$
    \hspace{1em}
    $\mathit{50.04}\,{\scriptscriptstyle\pm}\,\mathit{0.28}$}
  & \multicolumn{2}{c}{---} \\
\midrule
0
  & $\mathbf{20.49}\,{\scriptscriptstyle\pm}\,\mathbf{1.00}$
  & $\mathbf{48.75}\,{\scriptscriptstyle\pm}\,\mathbf{1.73}$
  & $10.00\,{\scriptscriptstyle\pm}\,0.00$
  & $49.60\,{\scriptscriptstyle\pm}\,0.72$ \\
3
  & $\mathbf{77.92}\,{\scriptscriptstyle\pm}\,\mathbf{0.47}$
  & $\mathbf{47.07}\,{\scriptscriptstyle\pm}\,\mathbf{0.37}$
  & $41.65\,{\scriptscriptstyle\pm}\,0.69$
  & $48.10\,{\scriptscriptstyle\pm}\,0.28$ \\
5
  & $\mathbf{83.14}\,{\scriptscriptstyle\pm}\,\mathbf{0.11}$
  & $\mathbf{47.49}\,{\scriptscriptstyle\pm}\,\mathbf{0.46}$
  & $46.08\,{\scriptscriptstyle\pm}\,0.41$
  & $48.24\,{\scriptscriptstyle\pm}\,0.66$ \\
10
  & $\mathbf{87.87}\,{\scriptscriptstyle\pm}\,\mathbf{0.17}$
  & $48.17\,{\scriptscriptstyle\pm}\,0.33$
  & $52.81\,{\scriptscriptstyle\pm}\,0.55$
  & $\mathbf{47.90}\,{\scriptscriptstyle\pm}\,\mathbf{0.43}$ \\
20
  & $\mathbf{90.73}\,{\scriptscriptstyle\pm}\,\mathbf{0.32}$
  & $50.14\,{\scriptscriptstyle\pm}\,0.81$
  & $60.70\,{\scriptscriptstyle\pm}\,0.65$
  & $\mathbf{47.57}\,{\scriptscriptstyle\pm}\,\mathbf{0.42}$ \\
30
  & $\mathbf{91.68}\,{\scriptscriptstyle\pm}\,\mathbf{0.12}$
  & $50.18\,{\scriptscriptstyle\pm}\,0.62$
  & $65.35\,{\scriptscriptstyle\pm}\,0.88$
  & $\mathbf{47.59}\,{\scriptscriptstyle\pm}\,\mathbf{0.48}$ \\
50
  & $\mathbf{92.61}\,{\scriptscriptstyle\pm}\,\mathbf{0.26}$
  & $50.21\,{\scriptscriptstyle\pm}\,0.41$
  & $70.92\,{\scriptscriptstyle\pm}\,0.62$
  & $\mathbf{47.29}\,{\scriptscriptstyle\pm}\,\mathbf{0.43}$ \\
\bottomrule
\end{tabular}}
\end{table}

\paragraph{Impact of forget ratio.}
Table~\ref{tab:sensitivity_forget_ratio} reports the effect of the forget ratio on
post-unlearning utility, MIA AUC, and unlearning time for the Adult data set with MLP.
Both variants of EUPG showed greater stability than Retrain.
Certified-SP was similarly stable, while SISA remained competitive in utility but much slower.
MIA AUC stayed close to $50\%$ for EUPG, Certified-SP, and Retrain throughout,
indicating consistently strong forgetting.
Thus, increasing the forget ratio mainly affected utility and runtime rather than the
membership signal.

EUPG was also the fastest across all forget ratios and its unlearning time generally
decreased as the forget ratio grew. 
The reason is that the retain data to be considered in fine-tuning decreases as the forget ratio increases.
Certified-SP showed a similar but slightly slower trend, whereas Retrain and especially
SISA remained much more expensive.

Overall, EUPG scaled favorably with the forget ratio, combining stable utility, strong
forgetting, and low unlearning cost.

\begin{table*}[t!]
\centering
\caption{Effect of the forget ratio on unlearning quality and efficiency: Adult data set, MLP, mean~$\pm$~std over 5~runs. In italics, the retrain baseline.
Among the remaining methods, boldface denotes the best and underlining the second best per column.}
\label{tab:sensitivity_forget_ratio}
\setlength{\tabcolsep}{4pt}
\resizebox{0.7\textwidth}{!}{%
\begin{tabular}{c l ccc}
\toprule
Forget ratio & Method & Utility~$\uparrow$ (\%) & MIA~AUC~$\downarrow$ (\%) & Time~(s)~$\downarrow$ \\
\midrule
\multirow{5}{*}{1\%}
 & \emph{Retrain}
 & $\mathit{85.25}\,{\scriptscriptstyle\pm}\,\mathit{0.07}$
 & $\mathit{50.52}\,{\scriptscriptstyle\pm}\,\mathit{1.03}$
 & $\mathit{17.21}\,{\scriptscriptstyle\pm}\,\mathit{0.17}$ \\
 & SISA
 & $85.31\,{\scriptscriptstyle\pm}\,0.12$
 & $\mathbf{49.00}\,{\scriptscriptstyle\pm}\,\mathbf{1.78}$
 & $36.36\,{\scriptscriptstyle\pm}\,1.24$ \\
  & Certified-SP ($\epsilon$=1, $\delta$=1e-5)
 & $85.59\,{\scriptscriptstyle\pm}\,0.05$
 & $\underline{49.54\,{\scriptscriptstyle\pm}\,0.69}$
 & $1.48\,{\scriptscriptstyle\pm}\,0.04$ \\
 & EUPG ($k$=30)
 & $\mathbf{85.63}\,{\scriptscriptstyle\pm}\,\mathbf{0.14}$
 & $49.61\,{\scriptscriptstyle\pm}\,0.78$
 & $\mathbf{1.21}\,{\scriptscriptstyle\pm}\,\mathbf{0.07}$ \\
 & EUPG ($\epsilon$=1, $\delta$=0)
 & $\underline{85.62\,{\scriptscriptstyle\pm}\,0.11}$
 & $50.09\,{\scriptscriptstyle\pm}\,0.95$
 & $\underline{1.22\,{\scriptscriptstyle\pm}\,0.10}$ \\
\midrule
\multirow{5}{*}{5\%}
 & \emph{Retrain}
 & $\mathit{85.21}\,{\scriptscriptstyle\pm}\,\mathit{0.10}$
 & $\mathit{49.66}\,{\scriptscriptstyle\pm}\,\mathit{0.97}$
 & $\mathit{17.99}\,{\scriptscriptstyle\pm}\,\mathit{0.13}$ \\
 & SISA
 & $85.29\,{\scriptscriptstyle\pm}\,0.08$
 & $\mathbf{49.75}\,{\scriptscriptstyle\pm}\,\mathbf{0.66}$
 & $34.49\,{\scriptscriptstyle\pm}\,0.34$ \\
  & Certified-SP ($\epsilon$=1, $\delta$=1e-5)
 & $\mathbf{85.63}\,{\scriptscriptstyle\pm}\,\mathbf{0.04}$
 & $50.01\,{\scriptscriptstyle\pm}\,0.89$
 & $\underline{1.50\,{\scriptscriptstyle\pm}\,0.05}$ \\
 & EUPG ($k$=30)
 & $85.58\,{\scriptscriptstyle\pm}\,0.10$
 & $\underline{49.79\,{\scriptscriptstyle\pm}\,0.78}$
 & $\mathbf{1.24}\,{\scriptscriptstyle\pm}\,\mathbf{0.11}$ \\
 & EUPG ($\epsilon$=1, $\delta$=0)
 & $\underline{85.60\,{\scriptscriptstyle\pm}\,0.16}$
 & $49.91\,{\scriptscriptstyle\pm}\,0.83$
 & $\mathbf{1.24}\,{\scriptscriptstyle\pm}\,\mathbf{0.10}$ \\
\midrule
\multirow{5}{*}{10\%}
 & \emph{Retrain}
 & $\mathit{85.20}\,{\scriptscriptstyle\pm}\,\mathit{0.16}$
 & $\mathit{50.22}\,{\scriptscriptstyle\pm}\,\mathit{0.76}$
 & $\mathit{16.36}\,{\scriptscriptstyle\pm}\,\mathit{0.67}$ \\
 & SISA
 & $85.33\,{\scriptscriptstyle\pm}\,0.10$
 & $49.81\,{\scriptscriptstyle\pm}\,0.51$
 & $32.98\,{\scriptscriptstyle\pm}\,0.24$ \\
 & Certified-SP ($\epsilon$=1, $\delta$=1e-5)
 & $85.57\,{\scriptscriptstyle\pm}\,0.03$
 & $\underline{49.66\,{\scriptscriptstyle\pm}\,0.35}$
 & $1.47\,{\scriptscriptstyle\pm}\,0.08$ \\
 & EUPG ($k$=30)
 & $\underline{85.61\,{\scriptscriptstyle\pm}\,0.10}$
 & $49.76\,{\scriptscriptstyle\pm}\,0.81$
 & $\underline{1.09\,{\scriptscriptstyle\pm}\,0.04}$ \\
 & EUPG ($\epsilon$=1, $\delta$=0)
 & $\mathbf{85.65}\,{\scriptscriptstyle\pm}\,\mathbf{0.16}$
 & $\mathbf{49.56}\,{\scriptscriptstyle\pm}\,\mathbf{0.52}$
 & $\mathbf{1.08}\,{\scriptscriptstyle\pm}\,\mathbf{0.05}$ \\
\midrule
\multirow{5}{*}{25\%}
 & \emph{Retrain}
 & $\mathit{85.00}\,{\scriptscriptstyle\pm}\,\mathit{0.08}$
 & $\mathit{49.67}\,{\scriptscriptstyle\pm}\,\mathit{0.55}$
 & $\mathit{14.31}\,{\scriptscriptstyle\pm}\,\mathit{0.38}$ \\
 & SISA
 & $85.16\,{\scriptscriptstyle\pm}\,0.11$
 & $50.17\,{\scriptscriptstyle\pm}\,0.50$
 & $28.43\,{\scriptscriptstyle\pm}\,4.09$ \\
 & Certified-SP ($\epsilon$=1, $\delta$=1e-5)
 & $\underline{85.53\,{\scriptscriptstyle\pm}\,0.08}$
 & $49.96\,{\scriptscriptstyle\pm}\,0.35$
 & $1.35\,{\scriptscriptstyle\pm}\,0.05$ \\
 & EUPG ($k$=30)
 & $85.50\,{\scriptscriptstyle\pm}\,0.12$
 & $\mathbf{49.79}\,{\scriptscriptstyle\pm}\,\mathbf{0.37}$
 & $\mathbf{1.01}\,{\scriptscriptstyle\pm}\,\mathbf{0.06}$ \\
 & EUPG ($\epsilon$=1, $\delta$=0)
 & $\mathbf{85.57}\,{\scriptscriptstyle\pm}\,\mathbf{0.14}$
 & $\underline{49.80\,{\scriptscriptstyle\pm}\,0.11}$
 & $\underline{1.02\,{\scriptscriptstyle\pm}\,0.08}$ \\
\midrule
\multirow{5}{*}{50\%}
 & \emph{Retrain}
 & $\mathit{84.31}\,{\scriptscriptstyle\pm}\,\mathit{0.25}$
 & $\mathit{49.93}\,{\scriptscriptstyle\pm}\,\mathit{0.39}$
 & $\mathit{9.33}\,{\scriptscriptstyle\pm}\,\mathit{0.11}$ \\
 & SISA
 & $85.19\,{\scriptscriptstyle\pm}\,0.13$
 & $50.04\,{\scriptscriptstyle\pm}\,0.19$
 & $16.10\,{\scriptscriptstyle\pm}\,0.16$ \\
 & Certified-SP ($\epsilon$=1, $\delta$=1e-5)
 & $\underline{85.48\,{\scriptscriptstyle\pm}\,0.07}$
 & $\underline{49.74\,{\scriptscriptstyle\pm}\,0.20}$
 & $1.26\,{\scriptscriptstyle\pm}\,0.14$ \\
 & EUPG ($k$=30)
 & $85.46\,{\scriptscriptstyle\pm}\,0.11$
 & $\mathbf{49.71}\,{\scriptscriptstyle\pm}\,\mathbf{0.30}$
 & $\underline{0.79\,{\scriptscriptstyle\pm}\,0.02}$ \\
 & EUPG ($\epsilon$=1, $\delta$=0)
 & $\mathbf{85.50}\,{\scriptscriptstyle\pm}\,\mathbf{0.04}$
 & $49.78\,{\scriptscriptstyle\pm}\,0.31$
 & $\mathbf{0.77}\,{\scriptscriptstyle\pm}\,\mathbf{0.02}$ \\
\midrule
\multirow{5}{*}{75\%}
 & \emph{Retrain}
 & $\mathit{83.39}\,{\scriptscriptstyle\pm}\,\mathit{0.15}$
 & $\mathit{49.80}\,{\scriptscriptstyle\pm}\,\mathit{0.36}$
 & $\mathit{5.28}\,{\scriptscriptstyle\pm}\,\mathit{0.09}$ \\
 & SISA
 & $\underline{85.23\,{\scriptscriptstyle\pm}\,0.20}$
 & $50.06\,{\scriptscriptstyle\pm}\,0.18$
 & $6.88\,{\scriptscriptstyle\pm}\,0.06$ \\
 & Certified-SP ($\epsilon$=1, $\delta$=1e-5)
 & $\mathbf{85.27}\,{\scriptscriptstyle\pm}\,\mathbf{0.18}$
 & $50.02\,{\scriptscriptstyle\pm}\,0.15$
 & $0.98\,{\scriptscriptstyle\pm}\,0.10$ \\
 & EUPG ($k$=30)
 & $85.14\,{\scriptscriptstyle\pm}\,0.13$
 & $\underline{50.00\,{\scriptscriptstyle\pm}\,0.12}$
 & $\mathbf{0.53}\,{\scriptscriptstyle\pm}\,\mathbf{0.02}$ \\
 & EUPG ($\epsilon$=1, $\delta$=0)
 & $85.19\,{\scriptscriptstyle\pm}\,0.11$
 & $\mathbf{49.93}\,{\scriptscriptstyle\pm}\,\mathbf{0.13}$
 & $\underline{0.59\,{\scriptscriptstyle\pm}\,0.06}$ \\
\midrule
\multirow{5}{*}{90\%}
 & \emph{Retrain}
 & $\mathit{82.61}\,{\scriptscriptstyle\pm}\,\mathit{0.54}$
 & $\mathit{49.85}\,{\scriptscriptstyle\pm}\,\mathit{0.29}$
 & $\mathit{2.92}\,{\scriptscriptstyle\pm}\,\mathit{0.09}$ \\
 & SISA
 & $\underline{84.87\,{\scriptscriptstyle\pm}\,0.13}$
 & $\mathbf{49.72}\,{\scriptscriptstyle\pm}\,\mathbf{0.10}$
 & $4.00\,{\scriptscriptstyle\pm}\,0.11$ \\
 & Certified-SP ($\epsilon$=1, $\delta$=1e-5)
 & $\mathbf{84.97}\,{\scriptscriptstyle\pm}\,\mathbf{0.14}$
 & $\underline{49.95\,{\scriptscriptstyle\pm}\,0.35}$
 & $0.88\,{\scriptscriptstyle\pm}\,0.13$ \\
 & EUPG ($k$=30)
 & $84.76\,{\scriptscriptstyle\pm}\,0.23$
 & $50.02\,{\scriptscriptstyle\pm}\,0.24$
 & $\underline{0.46\,{\scriptscriptstyle\pm}\,0.08}$ \\
 & EUPG ($\epsilon$=1, $\delta$=0)
 & $84.84\,{\scriptscriptstyle\pm}\,0.16$
 & $50.04\,{\scriptscriptstyle\pm}\,0.25$
 & $\mathbf{0.39}\,{\scriptscriptstyle\pm}\,\mathbf{0.04}$ \\
\bottomrule
\end{tabular}}
\end{table*}

\paragraph{Detailed runtime.}
Table~\ref{tab:efficiency_phases} separates one-time costs 
---data pre-protection (Prep) and pre-training--- from the fine-tuning (FT) on $\mathbf{D}$ and the unlearning cost (FT on $\mathbf{D}_r$).

It is quite clear that the dominant cost was pre-training the protected base model on the protected data, whereas unlearning only required a short fine-tuning time.
On all data sets, preparation and pre-training together accounted for most of the runtime.

EUPG incurred a larger upfront one-time cost, but this cost is more than amortized across subsequent forgetting requests, since each unlearning process reduced to fine-tuning on the retained data alone, taking about \(1.2\) to \(24\) s (Table~\ref{tab:efficiency_headtohead}).

\begin{table}[t!]
\centering
\small
\caption{Runtime breakdown of EUPG: forget ratio~$=5\%$, mean~$\pm$~std over 5 runs.}
\label{tab:efficiency_phases}
\setlength{\tabcolsep}{4pt}
\begin{tabular}{l c c c c}
\toprule
Data set & Prep~(s) & Pre-train~(s) & FT on $\mathbf{D}$~(s) & FT on $\mathbf{D}_r$~(s) \\
\midrule
\multicolumn{5}{l}{\textit{EUPG ($k$=30)}} \\
\midrule
Adult   & $8.45 {\scriptscriptstyle\pm} 0.47$   & $19.18 {\scriptscriptstyle\pm} 0.63$ & $1.20 {\scriptscriptstyle\pm} 0.09$ & $1.24 {\scriptscriptstyle\pm} 0.11$ \\
Heart   & $4.37 {\scriptscriptstyle\pm} 0.09$   & $22.61 {\scriptscriptstyle\pm} 1.26$ & $1.93 {\scriptscriptstyle\pm} 0.24$ & $1.82 {\scriptscriptstyle\pm} 0.08$ \\
Credit  & $13.42 {\scriptscriptstyle\pm} 0.26$  & $67.37 {\scriptscriptstyle\pm} 1.86$ & $5.99 {\scriptscriptstyle\pm} 0.30$ & $5.59 {\scriptscriptstyle\pm} 0.19$ \\
CIFAR10 & $326.51 {\scriptscriptstyle\pm} 6.21$ & $474.15 {\scriptscriptstyle\pm} 6.96$ & $24.71 {\scriptscriptstyle\pm} 0.53$ & $23.62 {\scriptscriptstyle\pm} 0.55$ \\
\addlinespace[2pt]
\multicolumn{5}{l}{\textit{EUPG ($\epsilon$=1, $\delta$=0)}} \\
\midrule
Adult   & $2.48 {\scriptscriptstyle\pm} 0.17$   & $9.48 {\scriptscriptstyle\pm} 0.60$  & $1.25 {\scriptscriptstyle\pm} 0.16$ & $1.24 {\scriptscriptstyle\pm} 0.10$ \\
Heart   & $4.35 {\scriptscriptstyle\pm} 0.21$   & $21.06 {\scriptscriptstyle\pm} 0.71$ & $1.87 {\scriptscriptstyle\pm} 0.17$ & $1.80 {\scriptscriptstyle\pm} 0.08$ \\
Credit  & $6.74 {\scriptscriptstyle\pm} 0.35$   & $66.82 {\scriptscriptstyle\pm} 2.27$ & $5.71 {\scriptscriptstyle\pm} 0.09$ & $5.55 {\scriptscriptstyle\pm} 0.25$ \\
CIFAR10 & $52.49 {\scriptscriptstyle\pm} 1.67$  & $473.55 {\scriptscriptstyle\pm} 6.54$ & $25.14 {\scriptscriptstyle\pm} 0.35$ & $23.53 {\scriptscriptstyle\pm} 0.09$ \\
\bottomrule
\end{tabular}
\end{table}

\section{Conclusions and future work}
\label{sec:conclusion}

We introduced EUPG, a model-agnostic framework for efficient machine unlearning with formal privacy guarantees.
EUPG protects the training data before model training and then handles forgetting requests by rolling back to a protected base model and fine-tuning only on the retained data.
Because the guarantee is tied to the protected data and inherited by post-processing, EUPG is compatible with arbitrary downstream ML models and with privacy guarantees beyond DP, including probabilistic \(k\)-anonymity. 

Experiments on tabular and image data sets with three ML models show that EUPG achieves the best trade-off between utility, efficiency, and unlearning effectiveness.
Compared with exact unlearning, it greatly reduces the computational and storage cost of repeated unlearning requests.
Compared with state-of-the-art certified unlearning, it applies to a wider class of models and privacy guarantees.
These results suggest that data pre-protection is more flexible and performs better than  in-protection certified unlearning methods that operate in the parameter space.
In fact, providing unlearning guarantees at the training data level as we do is more transparent and unequivocal than providing them at the level of the model parameters, since the same trained model or very similar trained models can be obtained from different training data.

Future work includes 
exploring hybrid designs that combine pre-protection with in-protection (in the parameter space); developing more utility-preserving formal protection mechanisms, especially for images and other high-dimensional data; and extending EUPG to broader ML settings, including sequence learning, large language models, diffusion models, and other model families relevant to unlearning. 
In fact, generative AI raises additional needs for data forgetting, such as copyright protection: the owner of copyrighted content can request its unlearning by an ML model, to prevent such content from being reproduced verbatim by the model. Our pre-protection approach can deal with this situation, by modifying copyrighted content so that it no longer appears verbatim in the training data.
Note that the vast size of the training data used in generative ML does not preclude using pre-protection: most of those training data are public-domain web data and their unlearning is unlikely to be requested, which means that the privacy guarantees only need to be enforced on the (small) fraction of non-public training data.

\section*{Acknowledgments}
Partial support to this work has been received from the Government of Catalonia (ICREA Acad\`emia Prizes to J. Domingo-Ferrer and to D. S\'anchez), MCIN/AEI under grant PID2024-157271NB-I00 ``CLEARING-IT'', and the European Commission under project HORIZON-101292277 ``SoBigData IP''.
We used Claude Sonnet 4.6 to optimize code implementation and WriteFull and ChatGPT-5.5 to correct typos, grammatical errors, and awkward phrasing throughout the article.

\bibliographystyle{IEEEtran}
\bibliography{my_bib.bib}

\end{document}